\documentclass[12pt,english]{article}
\usepackage{subfigure,tabularx}
\usepackage{mathptmx}
\usepackage{array}
\usepackage{bbding}
\usepackage{multirow}
\usepackage[fleqn]{amsmath}
\usepackage{amssymb}
\usepackage{amsthm}
\usepackage{graphicx}
\usepackage{natbib}
\usepackage{lscape}
\usepackage[hyphens]{url}
\usepackage[hidelinks]{hyperref}
\usepackage{comment}
\usepackage{bm}
\usepackage[title]{appendix}
\usepackage{longtable}
\usepackage{mathtools}
\usepackage{chngcntr}
\usepackage{authblk}
\usepackage{babel}
\usepackage{dsfont}
\usepackage{subcaption}
\usepackage{abstract}

\newcommand{\ADIM}{ADIM}
\newcommand{\SF}{SF}

\newcommand\norm[1]{\left\lVert#1\right\rVert}
\newcommand{\halvepagina}{0.5\textwidth}
\newcommand{\citeh}{\citep}
\newcommand{\citev}{\cite}
\newcommand{\fig}{figure~}
\newcommand{\eq}{equation~}
\newcommand{\Fig}{Figure~}
\newcommand{\tab}{table~}

\newcommand{\vct}{\overrightarrow}
\newcommand{\Rij}{R_{ij}}

\newcommand{\attractforce}{a}
\newcommand{\repforce}{F}

\newcommand{\maxdens}{\dens_\textrm{max}}
\newcommand{\ego}{\textrm{ego}}
\DeclareMathOperator*{\argmax}{arg\,max}
\DeclareMathOperator*{\argmin}{arg\,min}
\newcommand{\dens}{k}
\newcommand{\kcrit}{\dens_\textrm{crit}}
\newcommand{\qcrit}{\flow_\textrm{crit}}
\newcommand{\flow}{q}
\newcommand{\spd}{v}

\usepackage{amsmath}

\title{On Traffic Interactions for Unmanned Aerial Vehicles: Traffic Flow Applied to Three Dimensional Space}

\author[a,$\ast$]{Victor L. Knoop}
\author[a]{Serge P. Hoogendoorn}
\affil[a]{Delft University of Technology, Stevinweg 1, Delft, 2628CN, the Netherlands}

\def\abstract{Unmanned aerial vehicles (UAVs, or drones) are likely to significantly increase the amount of air traffic. If the skies are full of UAVs, they need to interact with each other, for instance by yielding or other evasive maneuvers. The aggregated movements of drones will create traffic patterns. 
Just like in current road traffic, the interactions will be very frequent, so a centralized computer managing these interactions is expected not to be possible. 
There is a long history of traffic flow theory and modeling for 1 dimensional (road) traffic; this has been expanded to 2 dimensional traffic (pedestrians). It is unclear how traffic flow theory works for 3 dimensional traffic. In this paper we show how drone traffic can interact in a decentralized way. For the microscopic description, we add asymmetric interaction rules.  We show that without centralized control, we can have efficient and safe traffic. 
Moreover, we provide a framework that directly links microscopic interactions to macroscopic properties. For the macroscopic description, we formulate and apply a numerical scheme that integrates the competition of space by UAVs for multiple classes, directions and dimensions. We apply both the microscopic and macroscopic descriptions to analyze (emerging) patterns which may arise in 3D traffic flow. 
The current paper provides background to develop interaction rules for drone traffic. Currently, the drone traffic is taking its first steps, but once the aeronautic technique takes off, the legislation regarding drone interactions should be ready. To support so, and be able to assess traffic consequences of decisions, the traffic flow theory framework developed here is essential.}

\begin{document}
\maketitle
\noindent \textbf{abstract}\newline \noindent \small{\abstract}

\section{Introduction}
It is a likely scenario that the amount of unmanned air vehicles (UAVs, drones) will grow significantly in the next decades, for instance for autonomous package delivery \citeh{doole2020estimation}. As of now, the way these UAVs will interact with each other is unknown. In the ideal future, they can all move at high speed close to each other in a straight line towards their destination. With higher densities, there will be relevant interactions, and UAVs will need to yield to other UAVs by changing their path and/or speed. This holds for any type of assumption one does for the interaction: at some point, the air space is simply too full to accommodate all UAVs.

As of now, no solid traffic flow theory for 3 dimensional (3D) traffic exists. Scientific theories on (1 dimensional) traffic streams go back for almost a century \citeh{Gre:1934}. Based on that, models and control have been developed. In parallel, the car usages have been expanded enormously, which in part has been made possible by the theories, models and controls. This step is yet to make for 3D traffic. There hence is a need to have a well established theory, models and control for air traffic. We will now first sketch the background available to develop these.

A first related field is car traffic theory. That has been described and modeled at different levels. The most common ones are the vehicle level (microscopic) describing the interactions between individual vehicles/drivers and the road level (macroscopic) describing the interactions between streams on roads. In the last decade and a half, also modeling and control on a zonal level (e.g., considering a neighborhood as one zone) has become increasingly popular \citeh{Dag:2007_1,mariotte2020calibration}. 

Pedestrian modeling is also related, taking the same elements as road traffic. Compared to road traffic it adds a dimension -- like air traffic --, yet it is less mature.  It has an empirically basis (e.g., \citev{Hel:2005}), where interactions are measured and modeled. These interactions can also be extended towards macroscopic properties like fundamental diagrams (e.g., \citev{hoogendoorn2017macroscopic}). Also dynamic models on a macroscopic scale have been presented, e.g., \citeh{Hoo:2014}. 

For air traffic, interactions are currently not as numerous as with cars. Moreover, they are (near busy areas like airports) controlled by a air traffic controller, following strategies which only are applicable for strongly centered traffic (towards a landing strip). Interaction strategies that UAVs can autonomously follow are not in use.

Recently, a couple of works have presented simulations to test whether  control on an aggregated level can also be used for air transport. 
The first efforts by \cite{Cum:2021} show that also in 3D traffic, a relationship between flow and density would emerge on the area level. \citev{tereshchenko2020macroscopic} show also MFDs for air traffic. They also mention effective distance covered, a measure of distance which only include the distance traveled in the shortest path. for non straight paths. The work by \citev{haddad2021traffic} utilizes the concept of an MFD to even in 3D apply perimeter control. Follow up works of e.g. \citeh{munishkin2023traffic} show that managing demand can optimize the traffic flow. A complete framework for managing demand to enhance performance is presented at ISTTT 25 \citeh{safadi2024integrated}.

The aim of this paper is to lie a further foundation for 3D traffic flow theory. We will argue that we can extend quite some concepts from road and pedestrian traffic to 3 dimensions.
We will explore potential microscopic interaction between UAVs based on instantaneous positions and speeds. Note that other concepts are also possible, most notably some type of ``road'' forming; these will be briefly discussed in the literature overview, but not analyzed in the paper. 

We will explicitly include breaking of symmetry in order to cater for higher speeds. These modeled microscopic interactions will determine the macroscopic patterns. Further elaboration of these interactions, and optimization thereof, will be the next steps in this domain. The current paper can be seen as a first theoretical step into this domain. The paper will derive some generic theories and concepts, which can be applied to UAVs of various shape and size; specific models and parameters need then to be updated accordingly.

This paper discusses UAV interactions on a microscopic scale and a macroscopic scale. After the literature review (section \ref{sec_literature_review}), the paper is split in a part on microscopic descriptions and macroscopic descriptions. We first handle microscopic descriptions, where we discuss the peculiarities of UAV traffic (section \ref{sec_requirements}) and then propose a (asymmetric) microcopic interaction model (section \ref{sec_models}). A case study (section \ref{sec_casestudy}) illustrates their working, and shows macroscopic patterns arising. The second part discussed macroscopic descriptions. We describe the principles and a macroscopic model (section \ref{sec_macroscopic}), and then present macroscopic case studies (section \ref{sec_macroscopic_case_studies}).  Finally, section \ref{sec_conclusions} will present the conclusions and the further outlook.  

\section{Literature review}\label{sec_literature_review}
UAVs are often seen as one of the modes of transport of the future. Consequently, the scientific field of transport has been going into this mode of transport in the recent past as well. The increased attention from the transport domain is related to the technical advancements in the UAVs themselves. We realize that these advancements enable and restrict the transportation field. However, that is a different scientific domain, and in this paper we focus on the traffic field of UAVs. Therefore, also studies that describe interactions for collective movements of UAVs as for instance a light show with UAVs at night,  are left out of this overview. We will explore two parts of traffic for UAVs: microscopic interactions (section \ref{sec_microscopicinteraction}) and traffic flow theory from a macroscopic point of view (section \ref{sec_macropscopic_interaction}).

\subsection{Microscopic interaction: models}\label{sec_microscopicinteraction}
Air traffic is typically controlled by an air traffic controller that requires two planes do not intrude each others area, requiring a minimum safe distance both horizontally and vertically. There are predetermined routes with planes with a similar speed and direction, leading to less conflicts. Currently, potential conflicts are typically solved by one of 4 strategies: (1) reduction of speed upstream and no change of path; (2) change of path of a plane towards another, predetermined path (3) circulation of a plane which wants to land in a suitable location; (4) change of path of a landing plane towards a landing such that its linear approach gets longer (i.e., merging further upstream). These approaches are developed for planes approaching landing areas. Only (1) is suitable for a generic location, yet it is unfeasible for busy traffic since might will lead to too low speeds. How UAVs should interact mid-air is hence not well established. 

That is not to say there are no interaction models being proposed. \citeh{viragh2016self} presented various interaction algorithms and compare optimize the parameters of the known interaction algorithms to have safe and efficient transportation. The algorithms used are mainly related to instantaneous and symmetrical interactions. The authors assess some macroscopic properties, and their main aim is investigate the effect of the models and the macroscopic efficiency. They explicitly include potential latency in communication, which could be a cause for degradation of traffic performance. The approach by \cite{soria2021predictive} illustrate the link between the UAV technique and the traffic operations: they developed algorithms to avoid obstacles and UAVs, and tested so, also using real drones. Their model uses predicted paths of other UAVs.

In another, preliminary work \citeh{hoogendoorn2023gametheoreticalapproachdecentralizedmultidrone}, we have considered the potential future conflicts of UAVs and how the trajectories can be optimized to minimize the conflicts. This is more in detail explained after the other used microscopic models, in section \ref{game_theory}.

At the other end of the spectrum, in the more traffic flow domain, \cite{bonnell2021velocity} proposed to improve 1-on-1 interaction on drones by directing traffic to the right (in their own frame of reference) if a conflict is detected. Simulations show that UAV traffic gets more efficient; note that the intervention is not essentially a 3D intervention -- a change to the right could work in 2D pedestrian traffic as well. Other studies (e.g., \cite{isufaj2022multi}), follow up and include not only 1-on-1 interactions, but aim to solve  conflicts by adapting the path of more drones. Modern machine learning tools can help solving these type of conflicts, without explicitly modeling them.

\cite{balazs2025decentralized} shows interaction algorithms for drones, their efficiency in simulation, and shows also the effect with in real live (up to 100 drones). They restrict their algorithm to a single flight level, and argue it can be coupled to various levels. That again is very much similar to pedestrian modeling, which is not seldom modeled in continuous 2D space, with interactions to different levels (at stairs or escalators). Note that UAV shows where 3D paths are being developed for show purposes (contrary to transportation purposes where a UAV needs to get to a certain destination) are a completely different field and not discussed here.

Based on the findings in literature, we want to expand the better established field of pedestrian modeling. Given the issues that might arise with communication, we stick to information which would -- in principle -- be observable by the UAV itself, and no communication (future flight plans, between-UAV negotiations) are needed. Moreover, we want to be ahead of potential UAV developments in developing theory, and hence do not restrict ourselves to what is feasible with current UAV techniques. Further developments in these techniques could be reason to further update the theories laid out here.

\subsection{Traffic flow theory}\label{sec_macropscopic_interaction}
There are various ways to organize air traffic. Apart from the centralized approach, there is a stream of researchers that advocate a decentralized approach \citeh{Hoe:1998}. This could be completely free, or UAVs can be pushed into predefined streams of UAVs with a similar heading and speed. This is like roads on the ground level, and work like levels or tubes in 3D space, for free areas airspace constraint by buildings: see e.g. \citeh{jang2017concepts, sunil2016influence, doole2020estimation, quan2021sky}. 

The interesting part is the link between the microscopic behavior and the macroscopic emerging patterns. The original definitions by \citeh{Edi:1965} can be applied to 3D traffic as well. Density $\dens$ is the number of UAVs per volume, or in generalized form, the sum of the time a UAV is present divided by an area in (3D) time and space (unit UAV/$m^3$). Flow $\flow$ is sum of the distances covered divided by an area in (3D) space and time (unit: UAV/$m^2/s$). Speed is defined as the ratio of flow and density 
\begin{equation}
\spd=\flow/\dens \label{eq_qisku}
\end{equation}

As mentioned, \cite{Cum:2021} paved the way and show that also in 3D traffic, a relationship between flow and density, using Edie's definitions, would emerge on the area level. MFDs  \citev{tereshchenko2020macroscopic} are also found in 3D traffic.  Empirical work \citeh{knoop2025macroscopic} shows that this already exist to some extent in plane current traffic. \citev{tereshchenko2020macroscopic} also exploit this for control, and they also mention effective distance covered, a measure of distance which only include the distance traveled in the shortest path. for non straight paths. The work by \citev{haddad2021traffic} utilizes the concept of an MFD to even in 3D apply perimeter control. Follow up works of e.g. \citev{munishkin2023traffic} show that managing demand can optimize the traffic flow. A complete framework for managing demand to enhance performance is presented at ISTTT 25 \citeh{safadi2024integrated}.

In pedestrian modeling, there are empirical and simulation works that discuss extensively the patterns and phenomena occurring. A very good and accessible overview is provided by \citev{Helbing2009}. It discusses the modeling, most notably the often used Social Force Model \citeh{Hel:1995}, which we will discuss extensively in section \ref{sec_models}.  Several phenomena are described, most interestingly ``stripes''\citeh{helbing2000freezing}. This means that there are clustered areas of pedestrians moving in the same direction, and in other areas there are pedestrians walking in other directions. In areas with traffic from different directions, these patterns are after the predictions empirically observed (e.g., \cite{mullick2022analysis}).

\part{Microscopic descriptions}

\section{Requirements for interaction models}\label{sec_requirements}
The interaction rules determine how UAVs interact with each other. There are several requirements, of which some are specific to UAV traffic. This section lists them.
\begin{itemize}
\item UAVs cannot come closer to each other than a predefined threshold for safety reasons.
\item Kinematic changes should satisfy physical restrictions of the UAVs. 
\item A minimum speed should be satisfied to prevent (some type of UAVs) from stalling, and a the trip length/duration should be maximized (the amount of energy to keep an UAV airborne is limited).
\item Movements should be smooth, to guarantee comfort/safety of the carried goods, but also to avoid rapid disruptions of the traffic state to which other UAVs can overreact.
\end{itemize}

The above list gives the requirements for the operations. 
Flight dynamics of UAVs suggest moreover that beyond the difference to how UAVs respond to objects in the front and in the rear (explicitly discussed and modeled), there are additional restrictions:
\begin{itemize}
\item Accelerating longitudinally requires thrust, and decelerating longitudinally requires a form of braking. This is a similar asymmetry as in car traffic. 
\item Changing direction goes via a different process than longitudinal acceleration. A similar asymmetry has been established for bicycles, which first have to change their steering bar, and then change direction \citeh{hoogendoorn2021game}. For winged UAVs it works even more indirect, via one additional layer. First, the flaps should be adapted. After this change of force, that UAV starts a roll/bank angle. Once banked, the wings provide the force required to make be in the banked turn. It takes the UAV additional time and distance to actually turn once banked.
\item Vertical accelerations are different in nature from acceleration in the horizontal plane due to gravity. Accelerating up (to higher levels) has more restrictions than accelerating down (to lower levels): which requires energy to be added when going up and not when going down.
\end{itemize}

\section{Mathematical description of microscopic models}\label{sec_models}
This section develops a model which can satisfy the criteria listed above. We start with a short recap of the (symmetrical) social force (SF) model (SFM) by \citeh{Hel:1995}, and then in section \ref{sec_ADIM} we present an asymmetrical addition which is better suited to the above criteria. Note the model is suited to the requirements and yielding smooth paths. Indeed, it will not be fully asymmetric in terms of dimensions, but we argue that with sufficient smoothness in the paths, it will also fulfill the more strict requirements for time to bank and going up. At the end, we also sketch a further model which goes beyond the analyses in this paper, which aims to increase safety by predicting the other UAVs paths, and even building in safety by risk-averse assumptions (section \ref{game_theory}).

\subsection{Social force model}
The SF model (SFM) combines 2 elements: the attraction towards his desired speed in case of no interactions, and the repulsion from the other UAVs. The force attracting a UAV to its desired speed is
\begin{equation}
\vct{\attractforce_i}=1/\tau*(\vct{v^0_i}-\vct{v_i})\label{eq_att_SFM}
\end{equation}
In this equation, $\vct{v^0_i}$ is the desired velocity of the UAV and $\tau$ is a time indicating a typical duration to reach that (note that it is not exactly the time at which the UAV will reach the speed).

At the same time, we have an repulsion of all other UAVs. In an equation, we obtain:
\begin{equation}
\vct{\repforce_i} = -A  \sum_{j \in \textrm{other UAVs}} \vec{n_{ij}} \exp \left(-\norm{\vct{\Rij}}/R0\right) 
\label{eq_rep_SFM}
\end{equation}
In this equation, $ \vec{n_{ij}}$ is the unit vector from i to j and $\Rij$ is the vector from i to j. 
For the parameters, we consider small-scale drones here and choose a free speed of 3.2 m/s, $\tau$=5s and $A$=2.1 m\textsuperscript{2}/s\textsuperscript{2}; $R0$ can be interpreted as minimum desired separation distance and is set to 20m.

These equations are all work in 3 dimensions. Moreover, they are fully symmetrical, so the repulsive force one UAV exerts on the other is equal to the repulsive force the other UAV exerts on the one UAV, like Newtons third law. This might not be realistic, as we will argue later.
The sum of forces \ref{eq_att_SFM} and \ref{eq_rep_SFM} acting on one UAV is the resulting force. 

\subsection{Asymmetric social force model}\label{sec_ADIM}
A known disadvantage of the above mentioned SFM is that interactions between two particles take place most once the particles become close to each other. Another effect is that particles that (in one cross-section, for instance at one altitude or ``flight level'', FL) move perpendicular to each other both repel each other, and instead of finding a place to cross, both move with each other (see also later an illustration in \fig \ref{fig_SF_trajectory_top_view}). 

To solve these issues, we modify the base SFM and add asymmetric interaction. We will refer to this as ADIM: Asymmetric Drone Interaction Model. The attraction term is fine, hence in the \ADIM, the attraction term is equal to the SFM, see \eq \ref{eq_att_SFM}.

\begin{figure}
\graphicspath{{figures/}}
\centering
\includegraphics[width=10cm]{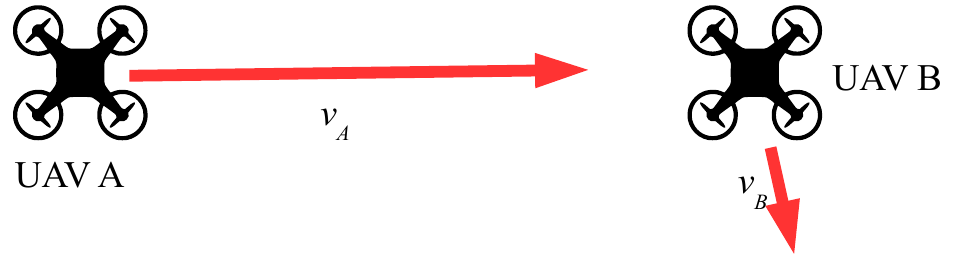}
\caption{The asymmetry in \ADIM}\label{fig_assymmetry}
\end{figure}

We follow the asymmetric version of the SFM, 
\begin{equation}
\vct{\repforce_i} = - A \sum_{j \in \textrm{other UAVs}} \vct{n_{ij}} \exp \left(-\norm{\vct{\Rij}}/R0\right) \left( \lambda_i \gamma_{ij} + (1-\lambda_i) \right)
\end{equation}
In this asymmetrical version, the interaction between $i$ and $j$ include a \emph{relative impact factor} $\gamma_{ij}$. Depending on the impact factor the repulsing force can be larger or smaller. The term $\lambda$ indicates to which extent the repulsive force is modified by the impact factor. A value $\lambda=0$ returns the SFM, whereas $\lambda=1$ takes the impact factor fully into account. This way, we can design the repulsion such that it varies \emph{depending on the relative velocities of 2 UAVs with respect to each other} and \emph{depending on the extent to which the UAV is flying towards the other UAV}. This is encapsulated in the impact factor. 

\subsubsection{Impact factor}
For this case, we define the impact factor itself in turn, by the relative speeds. It is first determined to which extent the UAV flies into the direction of the other UAV by the dot product of the (3D) velocity and relative speed vector, $\vct{Rij} \cdot \vct{v_i}$. If a UAV is flying away from another UAV, this dot product might become negative. However, a negative repulsive force (de facto an attractive force) would not make sense. Hence, we only take this term into account if positive, and if negative, we set the term to 0.

We also consider the speed: higher speeds will lead to a higher interaction term. Reasoning is that a higher speed will (at the same distance) lead to (a) a earlier conflict, and (b) a slower change of path with respect to distance (at the same acceleration). 
Therefore, we multiply this with the square of the speed, scaled with the reference speed $v^0$. In total, we hence define the impact factor $\gamma$ as
\begin{equation}
\gamma=\max\left(\vct{R_{ij}} \cdot \vct{v^0_i},0\right) \cdot  {\min \left(\frac{\norm{\vct{v_i}}}{\norm{\vct{v^0}}},1\right)}\label{eq_ADIN_IF}
\end{equation}
Combined, we find the collective repulsive force:
\begin{equation}
\vct{\repforce_i} =
 \sum_{j \in \textrm{other UAVs}}
 A \cdot \exp\left( \frac{-\norm{\vct{\Rij}}}{R0}\right) \cdot \left(\lambda \gamma_{ij} + (1-\lambda)\right)
\end{equation}
For the illustration, we choose $\lambda=0.9$ in the remainder of this paper unless stated differently. We do so since this keeps a small repulsive force even if the UAVs are not heading towards each other, and thus includes a additional safety margin.

Note that in this case the repulsive force of UAV A to UAV B is not opposite to the repulsive force of UAV A to UAV B. Since the speed of UAV is used in the equation of the impact factor of UAV A, it is now unique to the force acting on UAV A; see the illustration in \fig \ref{fig_assymmetry} In practice this means that a high speed UAV (in \fig \ref{fig_assymmetry} UAV A) that is approaching another UAV gets a higher force than a slow UAV (UAV B in \fig \ref{fig_assymmetry}) that happens to be in the way of an approaching UAV.

\subsection{Game theoretical negotiations}\label{game_theory}
The models above provide an interaction ``force'' between two drones. That force depends on the current position and speed of other drones. That means in practice information of the relative positions and, by considering the speed, a future position is known. However, speeds might change and there might be reasons for that. With other modes, we are used to anticipate on movement of other users, for instance cars braking and take a turn. In car traffic, we can provide models for vehicles navigating in 2 dimensions where drivers react to the planned paths of others \citeh{zhao2020two}, and similarly this holds for pedestrian movement. This can also be extended into a 3 dimensional movement.

In an earlier work \citeh{hoogendoorn2023gametheoreticalapproachdecentralizedmultidrone}, we have developed a game theoretical framework for interactions which we also propose for interactions between drones. Given safety issues, these might be relevant for air traffic, which is why we list them here, and indicate the relation to the instanteneous models we will use in the sequel.

In earlier work \citeh{hoogendoorn2013modeling, hoogendoorn2021game} it has been argued that there are 3 possible strategies to assume for a traveler:
\begin{itemize}
\item {Cooperation}: UAVs aim to jointy minimize joint cost of the interactions. This is a risk-prone strategy where a a UAV assumes the other UAV has the same strategy, helping him'
\item {Risk neutral} No collaboration or demon UAVs assumed: a user-equilibrium strategy where UAVs assumes that opponent behaves similar to the ego UAV, optimizing his path in his own interests. 
\item {Demon opponent} UAVs aim to minimize their own cost assuming that the other UAVs do aim to minimize the distance towards the ego UAV. This is a risk-averse solution where an ego drone assumes the worst to happen
\end{itemize}

Conceptually these strategies work as follows. An (ego, denoted $\ego$) UAV tries to consider what happens in the time span between now ($t_k$) and a time horizon further ($t_k+T$). He can control his action $u$. Then, the total amount of costs for a particular path ($J$) is the combination of its running costs $L$ during the period $t_k$ to $t_k+T$ and terminal costs $\Phi$ depending on the state at $t_k+T$. For the running costs, we consider a discount factor since future states are more uncertain and hence will get a lower weight:
\begin{equation}
J= \int_{t_k}^{t_k+T} e^{-\eta s}L(s,u,X) ds + e^{-\eta ({t_k+T})} \Phi\left(X(t_k+T)\right)
\end{equation}
Running costs typically consist of by its proximity to other UAVs, as well as control costs like changes in speed and altitude, and potentially other sources. The terminal costs indicate how attractive the state at the end of the time horizon is: it could be very attractive not to change path during the time horizon, but if that means that at the end an accident is unavoidable, that should be taken into account by a terminal cost. 

Now, we can consider three situations for the ego traveller A:
\begin{enumerate}
\item A and B both consider the costs for both and jointly minimize their combined costs: $\{u_A,u_B\}* = \argmin_{u_A,u_B}(J_A + J_B)$
\item A considers that B does not change it acceleration, and aims to minimize his own cost: $u_A* = \argmin_{u_A}(J_A|u_B=0)$ 
\item A considers the worst possible action of traveller B for the cost of A, and optimizes his own control function to minimize his travel cost:
$u_A*=\argmin_{u_A}\argmax_{u_B}J_A$
\end{enumerate}
These are even more relevant in air traffic, since traffic safety is more important. One can increase traffic safety by including a larger amount of distrust to other UAV. Consequence however, is that this will lead to less efficiency. 

Under specific choices for parameters, the game theory reverts to the social force model. If the discount factor $\eta$ is high, future costs will taken less into account and for the limit that $\eta$ approaches infinity, the future states will not contribute to the running costs. If the choice of $L$ then is matched to the SF model and final costs are set to zero, the approach simplifies to the SF model. 

\section{Case study}\label{sec_casestudy}
We test the model by simulating interactions between two clouds of UAVs.  

\subsection{Input}
\begin{table}
\caption{Complexity of interactions of cubes of n $\times$ n $\times$ n UAVs}\label{tab_interactions}
\begin{tabularx}{\textwidth}{Xccccc}
\hline
n&5&7&10&50&100\\
Number of UAVs \newline($2 \times n \times n \times n$)&250&686&2,000&250,000&2,000,000\\
Number of interactions \newline (number of UAVs $\times$ (nr of UAVs -1)) &62,250&469,910&3,998,000&6.2E10&4E12
\\\hline
\end{tabularx}
\end{table}

\begin{figure}
\graphicspath{{../figures/}}
\subfigure[Begin of interactions]{\includegraphics[width=\halvepagina]{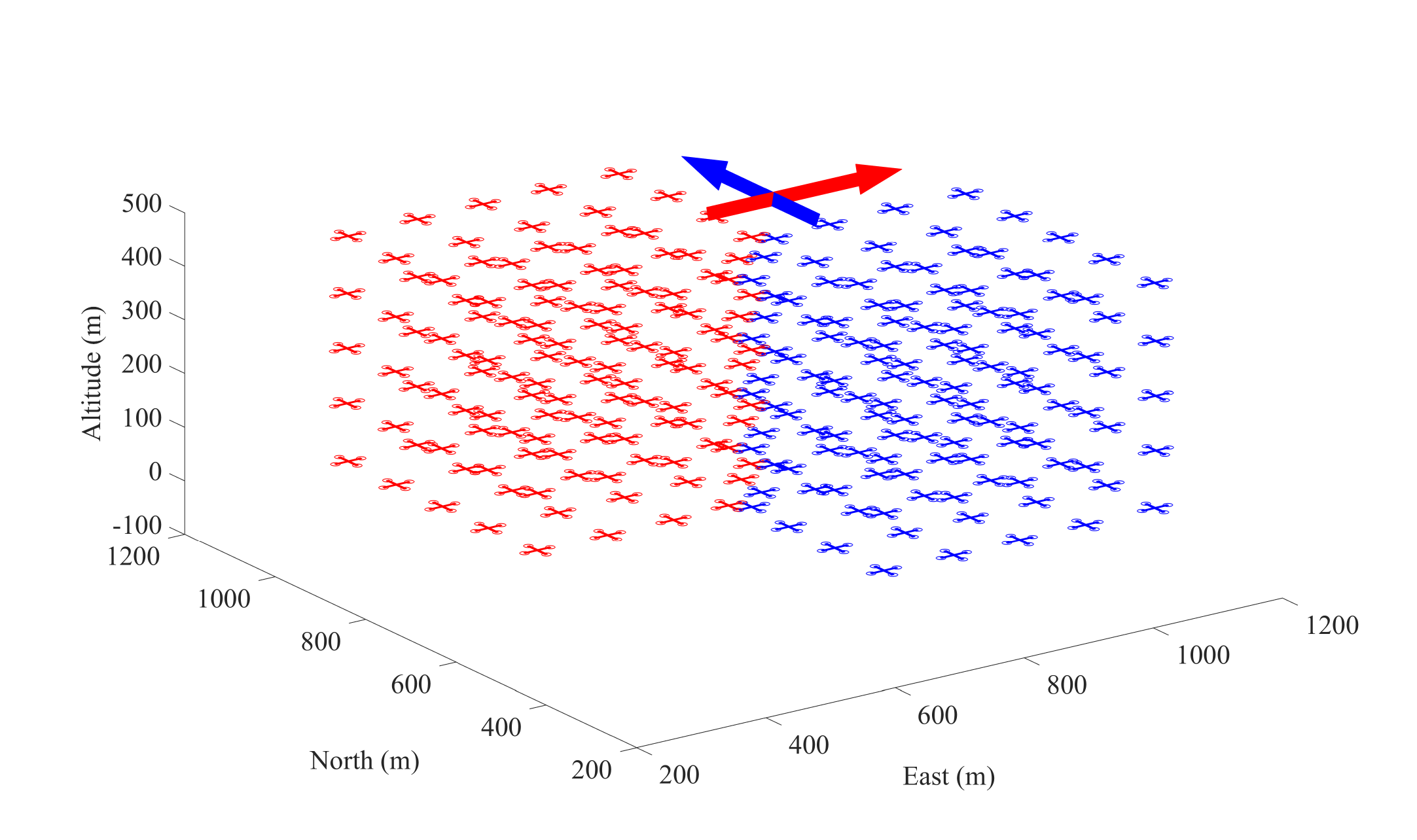}\label{fig_drones_start}}
\subfigure[UAVs passing each other]{\includegraphics[width=\halvepagina]{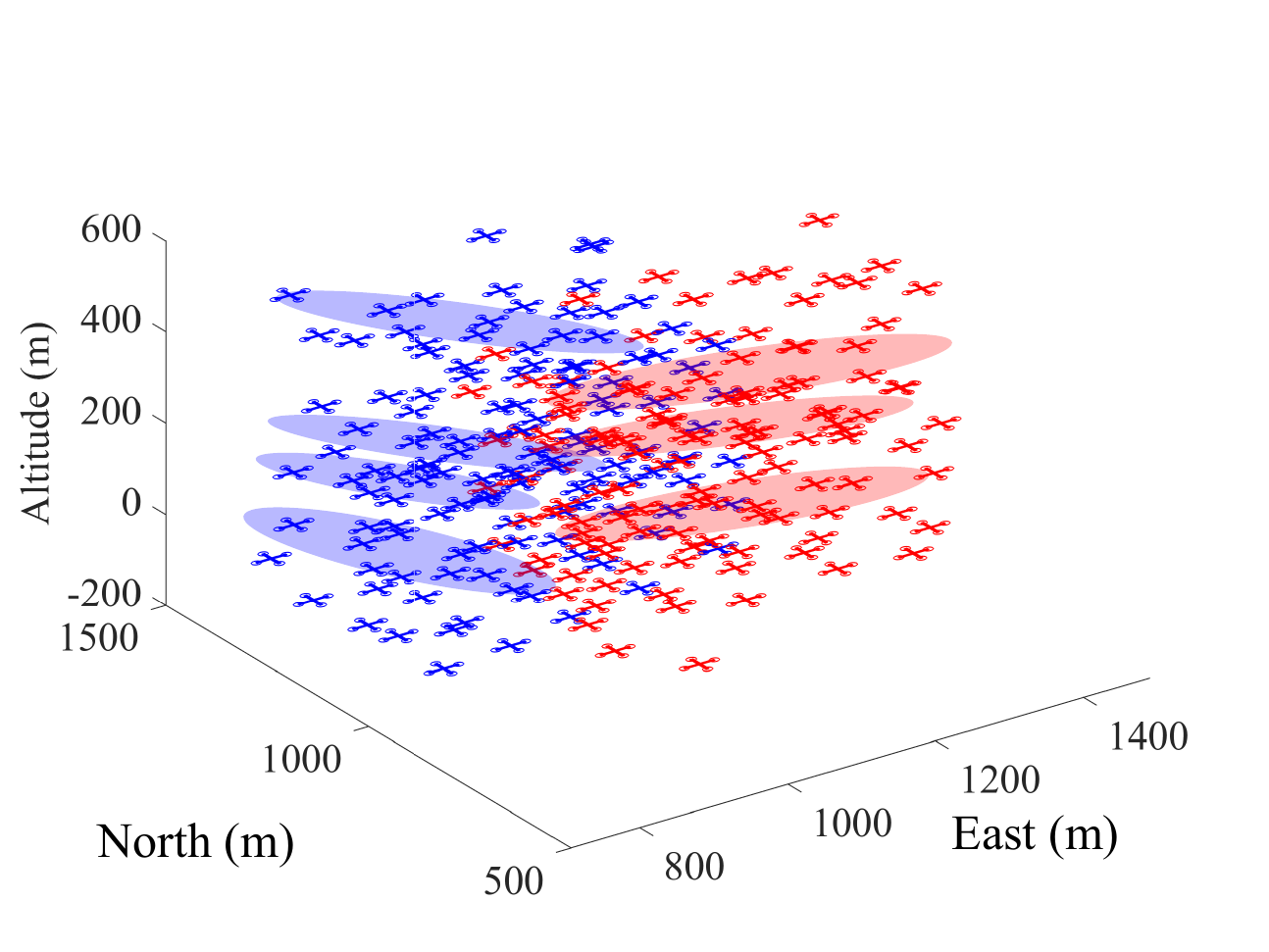}\label{fig_drones_interacting}}
\caption{The considered case study.}\label{UAVs_interacting}
\end{figure}
Two groups of UAVs each fly into a different direction (see \fig \ref{UAVs_interacting}) the red ones continue to the East(top right), and the blue ones continue to the North, which means they need to pass each other. We  tested the model for various numbers of UAVs. Note that the number and computational complexity rapidly increases. If we assume a cube of drones (size $n \times n \times n$),  table \ref{tab_interactions} shows the increasing complexity. Up to n=10 is easily feasible, yet scaling up to n=50 and n=100 is computationally more challenging. In this paper, we present the results for $5 \times 5 \times 5$ since there are sufficient UAVs to get emerging macroscopic patterns, and the lower number allows a better analysis and most importantly better visualization. 
 
For the \SF model, it is known that if two pedestrians walk straight towards each other, in the simulation, they will keep heading towards each other, since there is no lateral component of the force acting on each other. Hence they walk towards each other and stop, in principle forever. In order to avoid this avoid problems caused by numerical error, we slightly (millimeters) distort the position of the UAVs after every time step. Whereas the motivation to do so is numerically breaking the symmetry, one might realistically argue that wind might distort the drones, or relative positions are not well measured.

At this time step the first UAVs are getting in each others influence distance and have adapted their path (the grid is not regular anymore at the edge).
In particular, we start with two regular 3D grids of UAVs which fly into each other. Note we intentionally create different 3D clouds of UAVs to fully capture the 3D effect. All UAVs start at $v=0$, and aim towards their respective $v^0$. We use a grid of 5x5x5 UAVs (3 dimensions), all at regular intervals (100m grid spacing). Two groups of UAVs each fly along another axis, and they need to cross, see \fig \ref{fig_drones_start}. The minimum distance is set to 20m. 

Note they do not aim towards a destination but towards a directional velocity -- in our case along one of the axes. The difference is that after a distortion in position, they aim back to the same original directional velocity and they do not aim to correct for that distortion. We will simulate the standard SFM, as well as the asymmetrical model \ADIM.

\subsection{Measures of effectiveness}
The effectiveness of the interaction can be measured in various ways. In this paper, we consider three measures:
\begin{enumerate}
\item The location where UAVs start to deviate from the straight lines towards their destination, and how long the mutual interaction extends. This is an indication of the quality of the \nobreak{anticipation} behavior. Ideal anticipation starts early, and avoids UAVs making evasive maneuvers to the same side.
\item How much the UAVs deviate from their intended paths and speed. This is quantified by the lateral (i.e., orthogonal to the desired direction) deviation between the end position of the UAV and the original path (smaller is better). The same holds for travel time extensions (not reported in the extended abstract due to space limitations).
\item The closest separation to each other. This should not be below a safety threshold.
\item Accelerations for the UAVs. Ideally, they are as low as possible to ensure smooth paths, and paths which are within bounds of the physical restrictions for either direction. 
\end{enumerate}

We will also analyze the emerging properties, even ones that do not describe the effectiveness of the situation. For pedestrians, it is know that ``stipes'' occur: in a situation of crossing streams of pedestrians, those walking in the same direction will clog together. All properties in 3D traffic are hard to visualize, so we need a way to quantify. In this case, we choose the distribution of space headway to the 8 nearest UAVs (8 because that is the number of drones around in a 3 D grid space). At the start, the headways are all the same. If they clog together, one expects a large group of UAVs with a small headway, and a small group with a larger headway.

\graphicspath{{../figures/}}
\begin{figure}
\subfigure[Trajectories ADIM in 3D]{\includegraphics[width=\halvepagina]{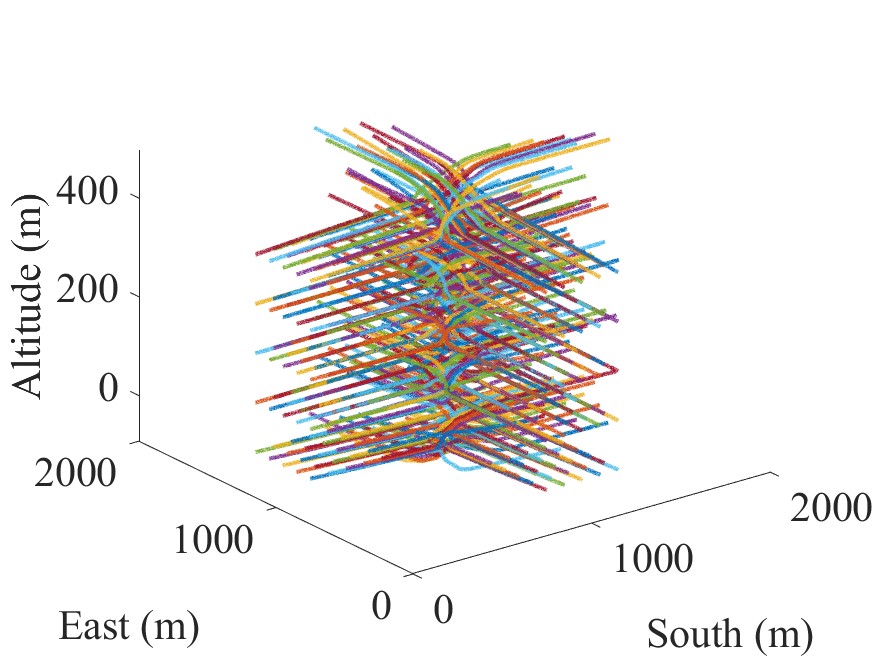}\label{fig_trajectory_view}}
\subfigure[Trajectories SFM]{\includegraphics[width=\halvepagina]{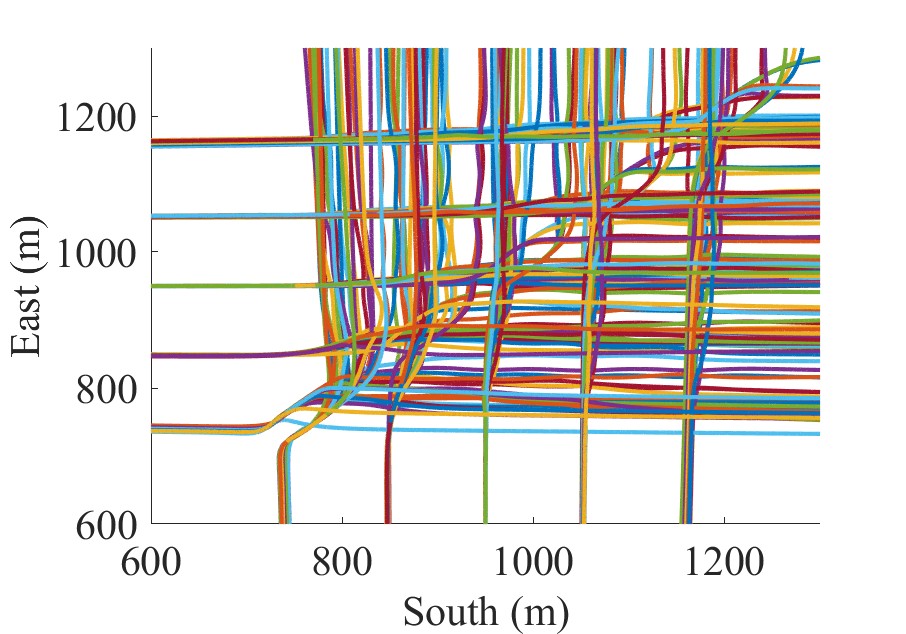}\label{fig_SF_trajectory_top_view}}\\
\subfigure[Trajectories \ADIM]{\includegraphics[width=\halvepagina]{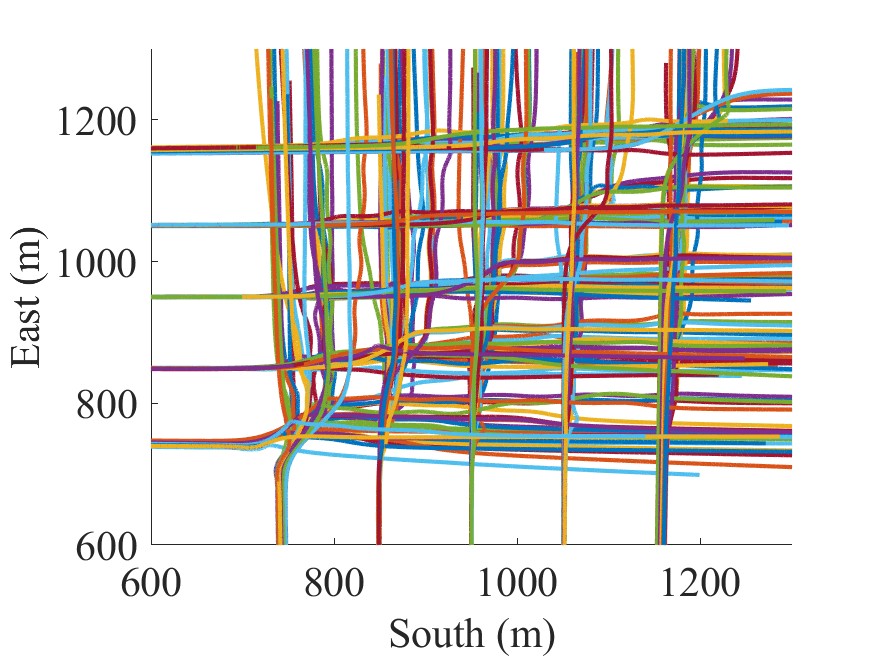}\label{fig_ADIM_trajectory_top_view}}
\subfigure[Lateral deviations of the paths (lateral to the intended direction)]{\includegraphics[width=\halvepagina]{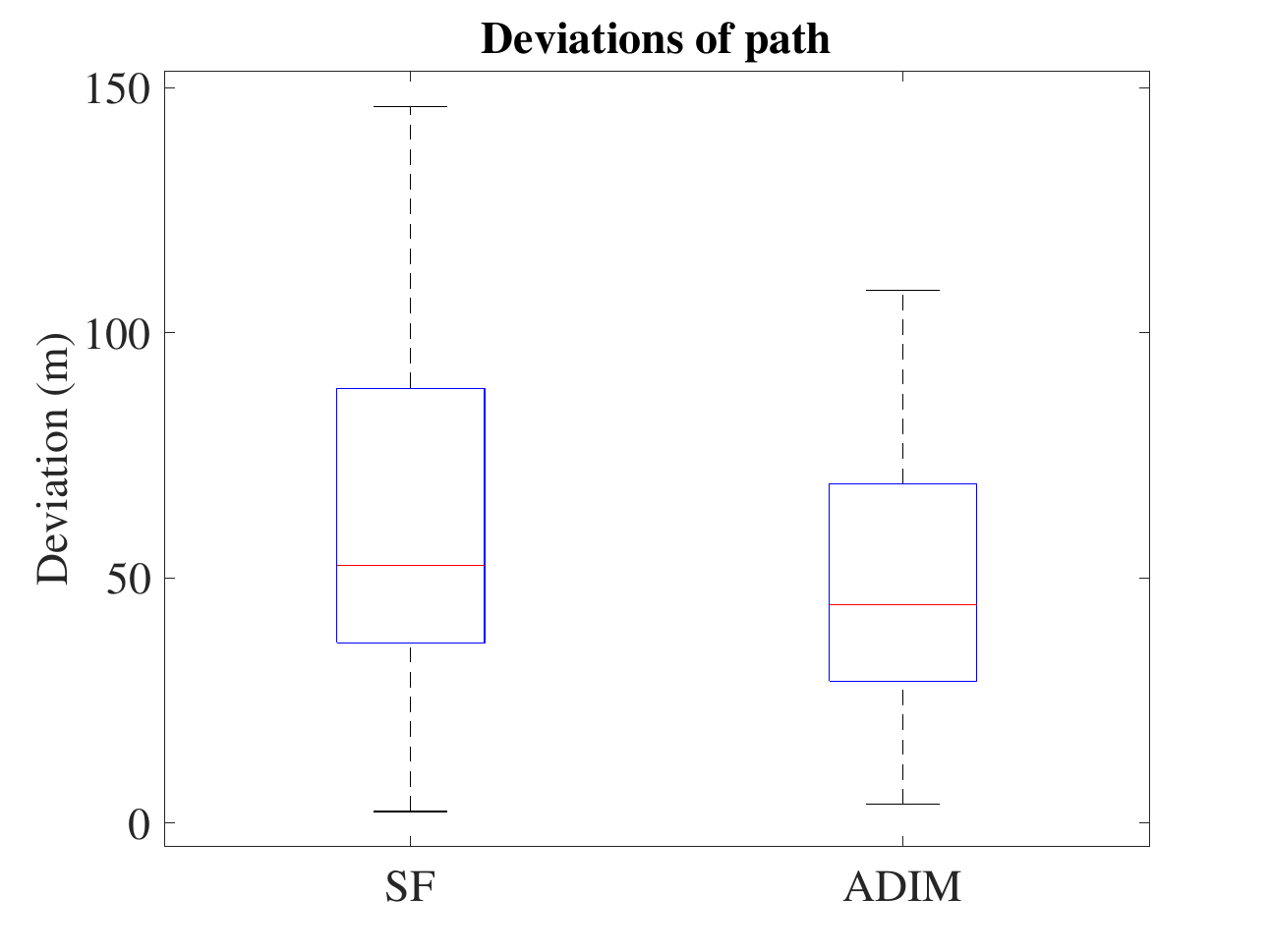}\label{fig_deviations}}\\
\subfigure[Smallest separation between any pair of UAVs]{\includegraphics[width=\halvepagina]{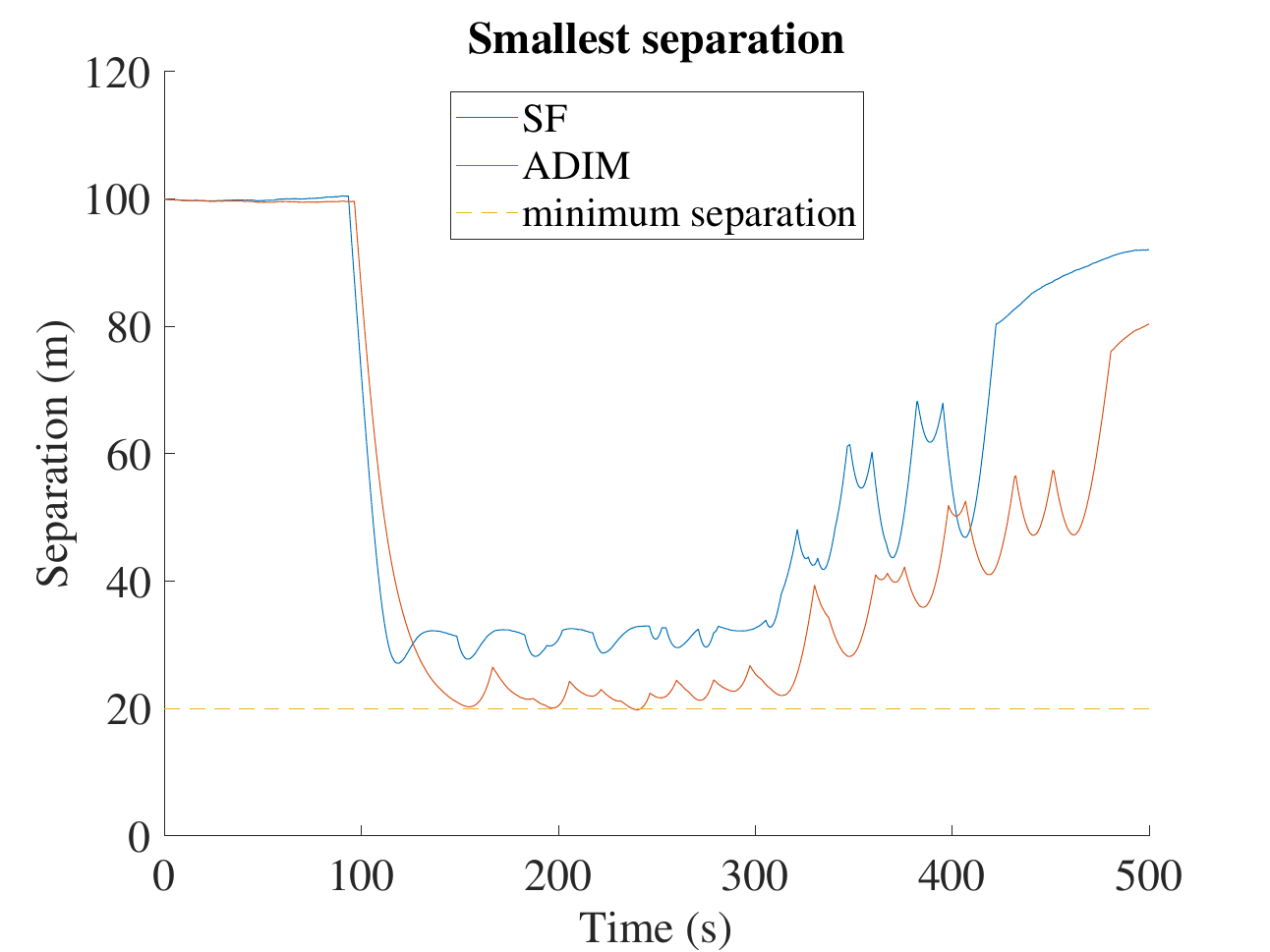}\label{fig_smallest_separation}}
\subfigure[Distribution of separations]{\includegraphics[width=\halvepagina]{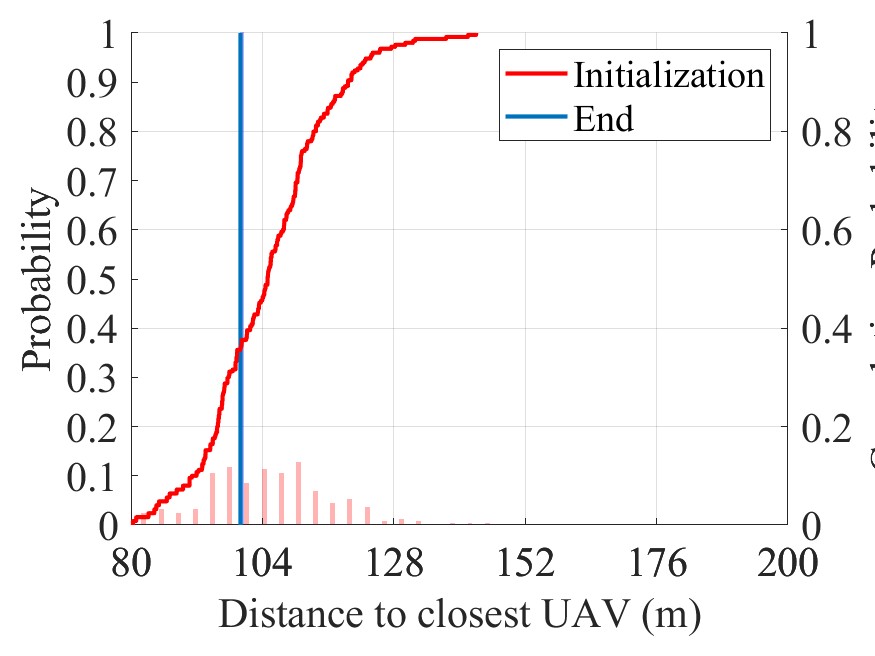}\label{fig_smallest_separation_distribution}}
\caption{Resulting trajectories}
\end{figure}


The operations are simulated at the microscopic level. We will construct fundamental diagrams relating it to the macroscopic level. We compute density and speed for every UAV. In order to get the density, we use a microscopic equivalent and compute the amount of which is closer to that UAV than to any other UAV, like in a Voronoi diagram, which we apply to 3 dimensions. The density related to a particular UAV is the inverse of the volume of its Voronoi cell. The speed is directly observable. Flow is, following \eqref{eq_qisku}, defined as density times speed. We compute so for all UAVs for all time steps, filtering out the cases related to the initial density.

\subsection{Results}
An impression of the resulting trajectories can be found in \fig \ref{fig_trajectory_view}. Since it is hard to analyse the way UAVs interact there, we show the trajectories also in the top view, in  \fig \ref{fig_SF_trajectory_top_view} and \ref{fig_ADIM_trajectory_top_view}. In the lower left corner of interactions with the SFM, we see UAVs in both directions moving diagonally; this is also somewhat visible with some UAV trajectories on the diagonal. The ADIM shows this to a lesser extent. Overall, the deviations of the intended path are generally less with ADIM than with the SFM, as is also shown in \fig \ref{fig_deviations}. 

\Fig \ref{fig_smallest_separation} shows the smallest separation distance between any pair of UAVs for each time step. At the start, the smallest separation is the initial separation. Then, once both ``clouds'' of UAVs approach each other, this reduces. It then goes up and down, with each time another pair of UAVs being the closest to each other. Overall we see that the separation is never below the set threshold of 20 meters ($R0$), for either model. We see that ADIM is able to follow this minimum closer, and is hence better in optimizing the available room to maneuver without violating the constraints. This hints that it might be more efficient in handling large groups of UAVs in a confined space. 

The clustering is shown by the distribution of the distance to the closest UAV, see \fig \ref{fig_smallest_separation_distribution}. This shows the result of 2 processes taking place at the same moment: a sprawl due to the fact that UAVs are repelling each other and at the same time they are moving close to each other when moving into the same direction. Moreover, we can visually identify clusters in the overview of the drones. Whereas quantitatively  this is hard to show, qualitatively the clustering is well visible in the produced movies. There is not a clear separation in the image, because there are layers, but in \fig \ref{fig_drones_interacting} we indicated with some ellipses some of the moving clusters. The clusters are longer in the direction of movement, indicating a platoon of UAVs which move in the ``void'' left (or made) by the platoon leader.

\begin{figure}
\subfigure[Density-speed]{\includegraphics[width=\halvepagina]{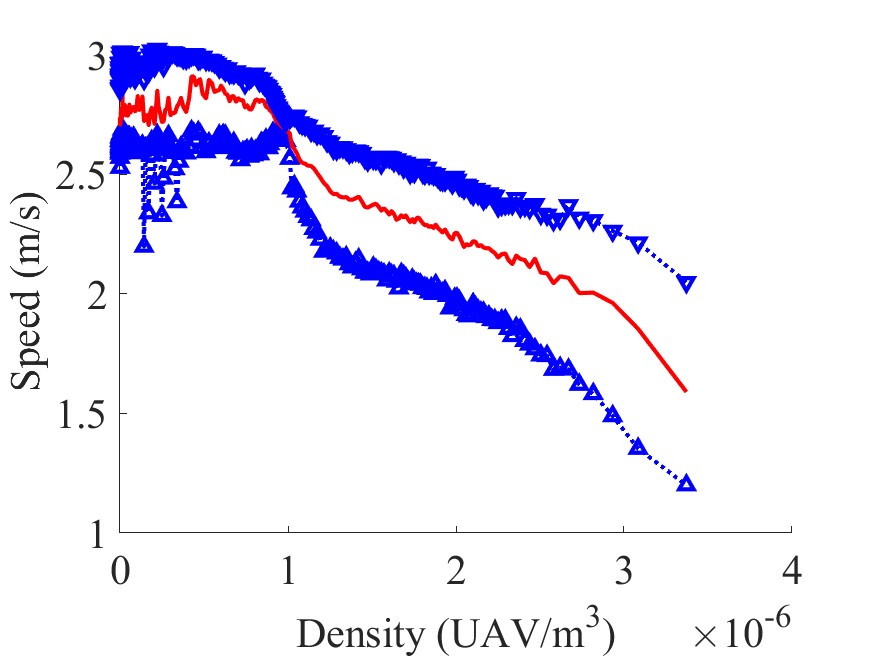}\label{fig_FD_vk}}
\subfigure[Flow-density]{\includegraphics[width=\halvepagina]{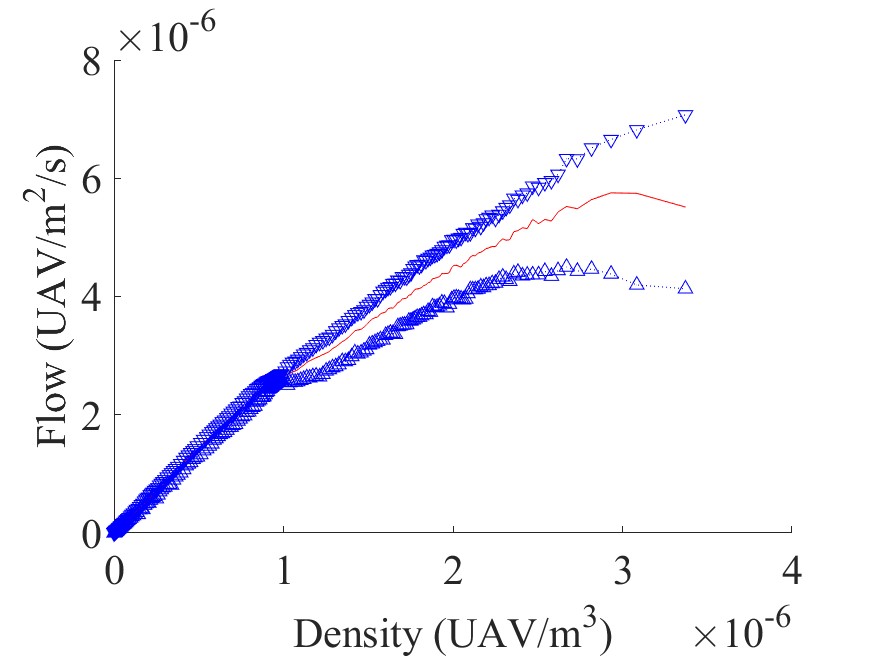}\label{fig_FD_qk}}
\caption{Fundamental diagrams}\label{fig_FDs}
\end{figure}
The fundamental diagrams \ref{fig_FDs} are as expected: we see first a constant speed which after a critical value starts to decrease. We also observe the resulting flow to flatten. The simulation conditions did not include heavy congestion, hence we do not see the congested part of the fundamental diagram in full.  

\part{Macroscopic descriptions}

\section{Macroscopic description}\label{sec_macroscopic}
If many UAVs will be present, a microscopic approach might be too detailed, for computational reasons as well as reasons of complexity of modelling. When upscaling to a macroscopic model, we argue that there is an average velocity $\vct{V}$ of UAVs in an area. This velocity can be class specific ($\vct{V_c}$). Amongst others, a different desired direction can define a different class of UAVs. We denote the velocities of all classes as $V_C$, which is a 3$\times n$ matrix, in which $n$ indicates the number of classes (and 3 the dimensionality of the space). Moreover, the density of the classes is class specific $K_C$, which is a 1$\times n$ vector. 

Let's split the space in cells.  A straightforward upscaling of any microscopic model to a macroscopic model is as follows. We can transform all equations of microscopic interaction into a macroscopic form. For each cell $i$, the class-specific density and the class-specific velocity is known. This gives the an interaction term between (all) the UAVs in the current cell and the other cell, by multiplying the individual interaction by the number of UAVs in the other cell. This can be used to adapt the aggregated velocity in the current cell. 
With a known velocity and and density, a new density can be formed. For this, a Godunov-like scheme can be used, like presented in \cite{Wag:2018}. This yields new speeds and flows, yielding in turn new densities. 

Note that computationally, this changes the number of interactions from the square of number of UAV to the square of the number of cells. This is hence useful for a low number of cells, or a high number of UAVs. For a high number of cells, this might still be unfeasible. Limiting the length of interactions could limit the computational complexity. We did so in the approach presented in the sequel, where we realize that interactions are local.

We can build a macroscopic model around the social force model in a generic form including an asymmetry factor, as our ADIM model. This has been done for a 2D model, see \citeh{Hoo:2014}. Exploiting symmetries and integrating the forces over all directions, we can derive equations for an equilibrium velocity per class. We can extend this analysis to three dimensions.

 
%
%
%
%
%
%

We can directly apply the derivation as mentioned by \citeh{Hoo:2014}, adapting the interpretations for 3 instead of 2 dimensions. In equations, the derivations remain identical. Based on the anisotropic SFM, we obtain the following multi-class equations:
\begin{eqnarray}
\norm{V_c} &=& 
\norm{\vct{v^0_c} - \sum_{\textrm{all classes c'}}
\norm{\beta_{c' \rightarrow c} \cdot \vct{\nabla K_{c'}}} }
- \sum_{\textrm{all classes c'}} \alpha_{c'\rightarrow c} \cdot K_{c'}\\
\vct{e_{Vc}}
&=&\frac{\vct{v^0_c} - \sum_{\textrm{all classes c'}}
\beta_{c' \rightarrow c} \cdot \vct{\nabla K_{c'}} }{\norm{\vct{v^0_c} - \sum_{\textrm{all classes c'}}
\beta_{c' \rightarrow c} \cdot \vct{\nabla K_{c'}}} }\label{eq_velocities}
\end{eqnarray}
In this, $\vct{e_{Vc}}$ indicates the direction of the velocity $V_c$. $\alpha$ and $\beta$ are given by
\begin{eqnarray}
\alpha_{c'\rightarrow c} &=& \pi \tau (1 + \lambda) A_{c'\rightarrow c} (R0)^2_{c'\rightarrow c}\\
\beta_{c' \rightarrow c} &=& 2 \pi \tau (1 - \lambda) A_{c'\rightarrow c}(R0)^3_{c'\rightarrow c}.
\end{eqnarray}

The elegance of this approach is that the effective change of speed only depends on the \emph{local} effects, i.e. the gradient in density for each of the classes of UAVs and the velocities of the different classes of UAVs. The need to compute interacting forces between all elements is eliminated.

The model produces non homogeneous traffic states even under steady input conditions. As argued in \citev{Hoo:2014} for pedestrians (with pedestrian-specific tuning), this model type can recreate spatial separation of the streams in various directions. Also for 3D traffic we expect the model to endogenously show this directional separation of streams, and to address the competition for space for the classes.

\subsection{Numerical model}
Compared to a single dimensional road where traffic flows in one direction, in this case there is traffic from various dimensions competing for the same space. This paper proposes to solve this issue in an innovative way, by integrating so in the numerical scheme. We choose a model that combines demand and supply for each cell, and in the numerical model allows for the various classes.
\begin{figure}
\includegraphics[width=\textwidth]{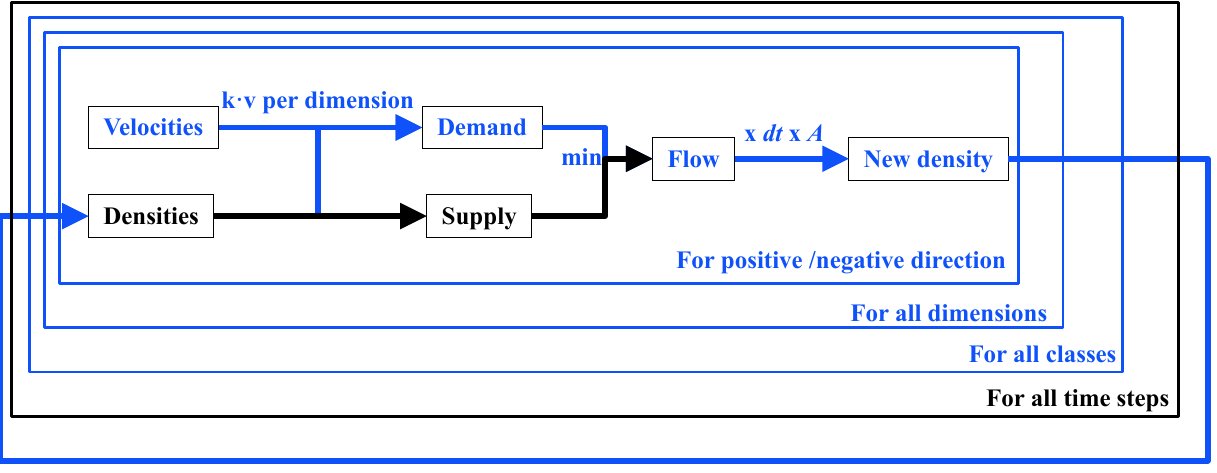}
\caption{The numerical scheme. Blue lines means it is subset-specific, and black lines means it is for all subsets combined/aggregated.}\label{fig_numerical_model}
\end{figure}
We follow the numerical model of \citeh{mollier2019two}, which is a 2D multi-class variant of a Godunov scheme with a demand and supply. In short, is works as follows (see also \fig \ref{fig_numerical_model}). Whereas the numerical  scheme is known, we are the first to exploit this to overcome the inherent issues of macroscopic modeling of 3D traffic. 

We split time into time steps, and then further divide the time steps in substeps. Within a time step, we iteratively compute the flow for \emph{one} subset. We define a subset as a specific (a) classes of drones (different destinations), flying into a specific (b)  dimensions (i.e., east-west, north-south, bottom-top), and a specific (c) directions for each of these dimensions (i.e., east or west; north or south, etc.).  We repeat that for all subsets, and once we have computed flows for all subsets (i.e., all classes, all dimensions and all directions), we move to the next subset. 

The cleverness lies in the fact that during the determination of the flow for a substep, we compute the flow which we assume to hold for the complete time step. The supply is determined by the actual aggregated density of all classes. The new density for that class is determined after the substep, and this new density is taken into account in determining the supply for the next subset. The competition for space by different classes, different dimensions and different directions is thus handled by the numerical scheme. 

We chose to keep the handling order subsets the same for all time steps. In theory this could imply that a cell is occupied up to critical density by traffic from one subset, just beating the other subset by a substep. This could be overcome by in each time step chose a random order for the subsets. Having said so, with demands changing at time scales slower than one time step and a typical fundamental diagram which cannot fill a cell to jam density within a time step (at maximum critical density is), this is not likely to be an issue. 

The demand is the dot product of the the velocity, and the considered direction (e.g., east), multiplied by the density. This is based on the continuity equation ($\flow=\dens \times \spd$, eq. \ref{eq_qisku})), following \cite{Edi:1965}. The velocity is determined by \eqref{eq_velocities}. We split the flow in opposing directions (i.e., we split east and west) and do not take negative flows into account. We do so because we do not want flows to cancel out; flow in opposite directions will be considered in different sub-timesteps. Once the flows at the boundary areas are known, we multiply by the surface area connecting the zones (perpendicular to the direction) and time step to get the flows in UAVs.

This means that we need to get the density (we keep track of the densities per cell), and find a speed. Recall that the speed is given by:
\begin{eqnarray}
\norm{V_c} &=& 
\norm{\vct{v^0_c} 
- \sum_{\textrm{all classes c'}}
\norm{\beta_{c' \rightarrow c} \cdot \vct{\nabla K_{c'}}} }
- \sum_{\textrm{all classes c'}} \alpha_{c'\rightarrow c} \cdot K_{c'}
\end{eqnarray}

Following principles of demand and supply (as also in \citeh{Dag:1994}), we differentiate between undercritical and overcritical conditions in the scheme. Demand and supply in overcritical and undercritical conditions are set at the capacity of the cell. Hence, we need an additional step to distinguish undercritical and overcritical conditions, and find the capacity. 
Since within a cell, we consider equilibrium conditions, hence $\nabla \rho$=0. So we have 
\begin{eqnarray}
\norm{V_c} &=& 
\norm{\vct{v^0_c}} - \sum_{\textrm{all classes c'}} \alpha_{c'\rightarrow c} \cdot K_{c'}
\end{eqnarray}
This is a Greenshields \citeh{Gre:1934} shape fundamental diagram where speed decreases linearly with density. The maximum density is found at $\maxdens=v_0/\alpha$ for which the maximum flow is found at $\kcrit=v_0/\alpha$ at the value:
\begin{equation}
\qcrit=\kcrit v=\kcrit v_0/2 = v_0^2 /2 \alpha
\end{equation}
In this approximation, we ignore multi-class corrections which might lower the critical density or the maximum flow. This might temporarily lead to overcritical conditions. However, that will be taken into account in the next computation, when the speed in the cell is being adapted to that higher density. In the implementation, we furthermore add two safeguards due to potential numerical errors: (1) the number of UAVs flowing out of a cell can not be larger than the number of UAVs present in that cell; (2) the speed is at least 0.1 m/s. The speed has no maximum, but should be bounded by the CFL condition \citeh{courant1967partial} stating UAVs should not travel more than one cell per time step. Since this condition is not met by default, we will check on this condition at every time step.

Apart from modeling the open sections, we might want to model restricted areas to fly, for instance high rise buildings. The model itself has no option to restrict the space to fly, so we propose and implement a method based on the model parameters. The restricted areas do not allow any UAVs to fly. In modeling, there are two options to ensure these will not be used: (a) reduce the supply for these models (via the numerical implementation); (b) in the determination of the speeds, take the increase the densities for the blocked cells such that they have a repelling effect of flows (via the modeling). Option (a) is effective to not allow any UAVs in and will be used for blockings; the addition of (b) could lead to an earlier deviation of paths. We will show the effects of these implementations in the case studies.

\section{Macroscopic case studies}\label{sec_macroscopic_case_studies}
We will first describe the setup of the case study (section \ref{sec_macroscopic_case_study_setup}), followed by the results (section \ref{sec_macroscopic_case_study_result}).
\subsection{Setup}\label{sec_macroscopic_case_study_setup}
The case study will show the traffic phenomena as the macroscopic model reveals. We will address path derivation in the same flight level (altitude), patterns of congestion, including severely congested cells (``gridlock'' within a cell), and analyze the effect of $\lambda$ and the effect of modeling the blocking.

We will illustrate the working of the model based on several scenarios, all with a different initialization in terms of UAV location and direction. In all cases we initialize the state with UAVs in the system; UAVs reaching the end of the system fly out of the system, and there are no inflowing UAVs.  In all cases, we break symmetry by initializing UAVs distributed over the different heights. An overview of the scenarios we carry out, is shown in \tab \ref{tab_macroscopic_scenarios}. 
\begin{table}
\caption{Scenarios for the macroscopic study}\label{tab_macroscopic_scenarios}
\begin{tabularx}{\textwidth}{lXlll}
\hline
Number & Directions & Density & altitude&anisotropy\\\hline
1 & east-west and west-east & medium & higher half&anisotropic\\
2 & left-right and front-rear & medium & higher half resp lower half&anisotropy\\
3 & left-right and front-rear & high & higher half resp lower half&anisotropy\\
4 & left-right and front-rear & high & higher half resp lower half&isotropy\\
5 & east-west and west-east\newline including blocking & lower & higher half&anisotropic\\
\hline
\end{tabularx}
\end{table}

The first scenario is analyze the working of the model and the fact that UAVs indeed move  around each other. For the second case study, we aim to have a scenario similar to the microscopic scenario, where two groups of UAVs approach each other under a 90 degree angle. Also here, we break symmetry by having a different distribution of UAVs over various layers. 
We repeat that with various parameters: higher densities (scenario 3), and a different anisotropy value ($\lambda$, scenario 4). Scenario 5 shows the effects of blockings, which we will do with a restriction of supply only (5a) or with an added density for speed field calculations as well (5b).

\graphicspath{{../macroscopic/intermittant_class_and_dimension_time_steps/figures/}}
\begin{figure}
\includegraphics[width=\textwidth]{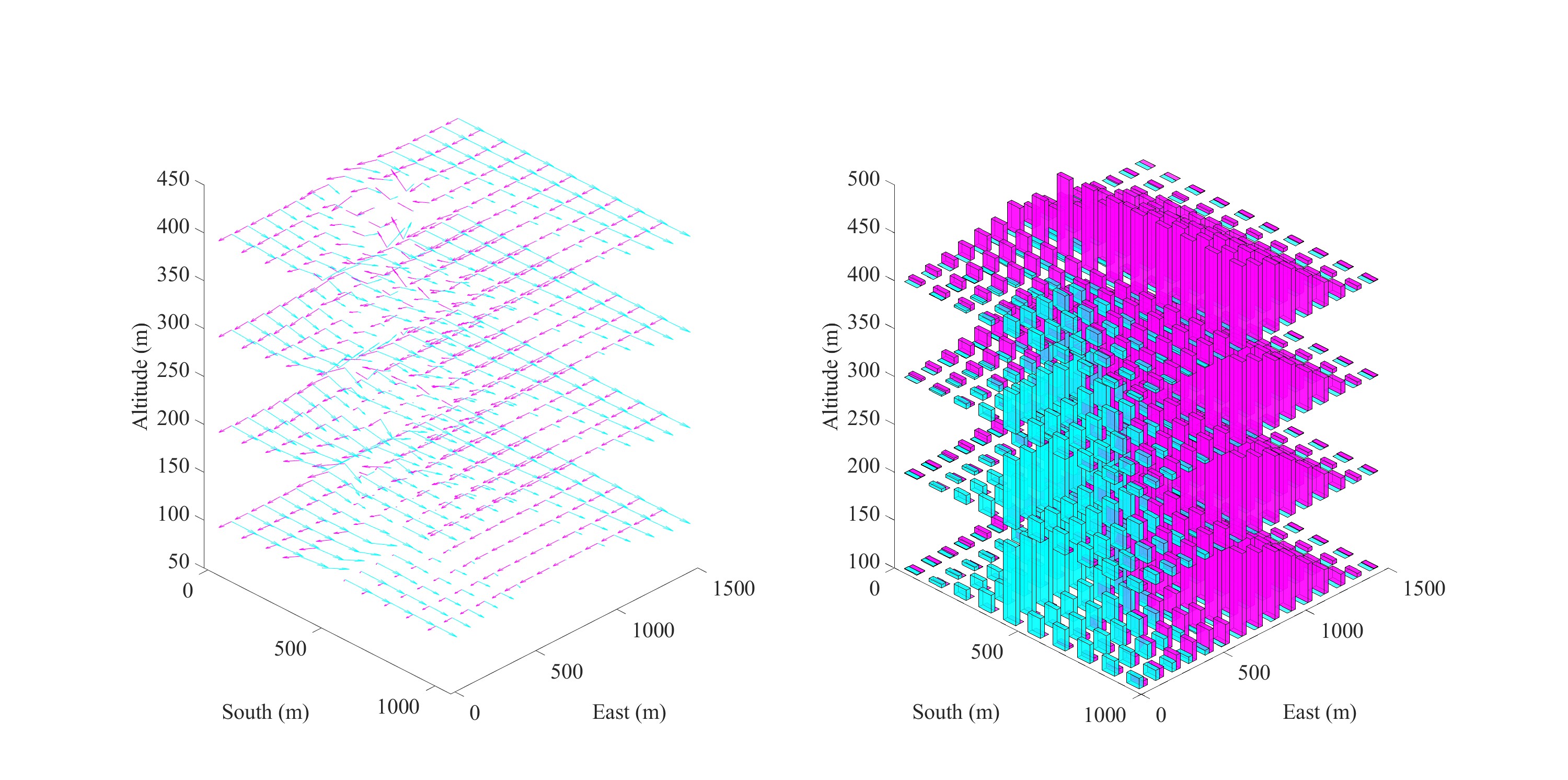}
\caption{Velocity field (left) and density for a 3 dimensional space; 2 different colors indicate different classes}\label{fig_density_and_speed}
\end{figure}
The analysis of the macrocsopic model is complex. We have a 3 dimensional grid of cells. In all cells, there are for each class a density of drones, as well as a 3-dimensional speed vector. This is visually shown in \fig \ref{fig_density_and_speed}. Whereas it is possible to see, we will for the squeal of the paper opt to show 2-dimensional graphs, and comment on the third dimension, see e.g. \fig \ref{fig_init_opposing}. These will show the density (size of the bar) for different classes (colors) and the 2D speed in each of the cells for each class (arrows in different colors); only the vertical speed is not visible in this representation. We will represent that by altitude distributions.

\subsection{Results}\label{sec_macroscopic_case_study_result}
First, we 
\begin{figure}
\subfigure[Level operations]{\includegraphics[width=\halvepagina]{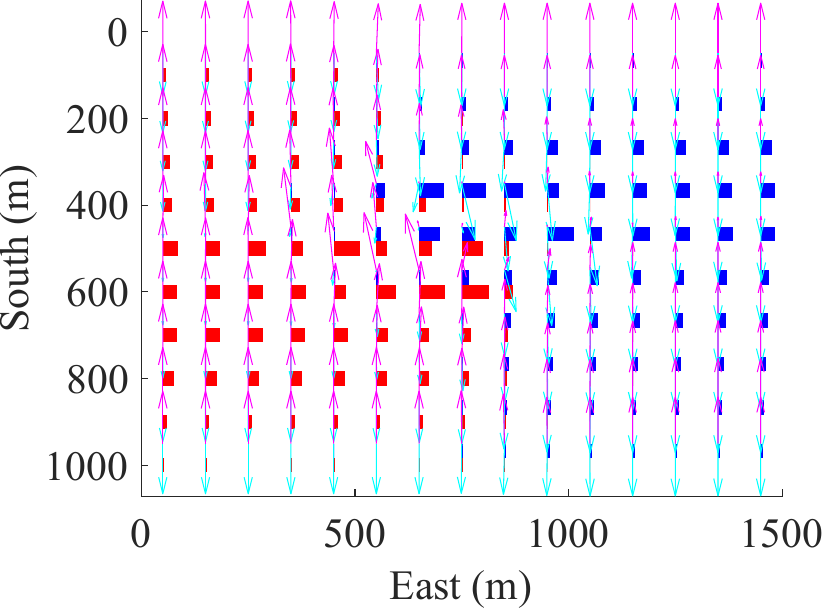}
\label{fig_opposing_directions_inFL}\label{fig_init_opposing}}
\subfigure[Vertical distribution; different line styles indicate different classes]{\includegraphics[width=\halvepagina]{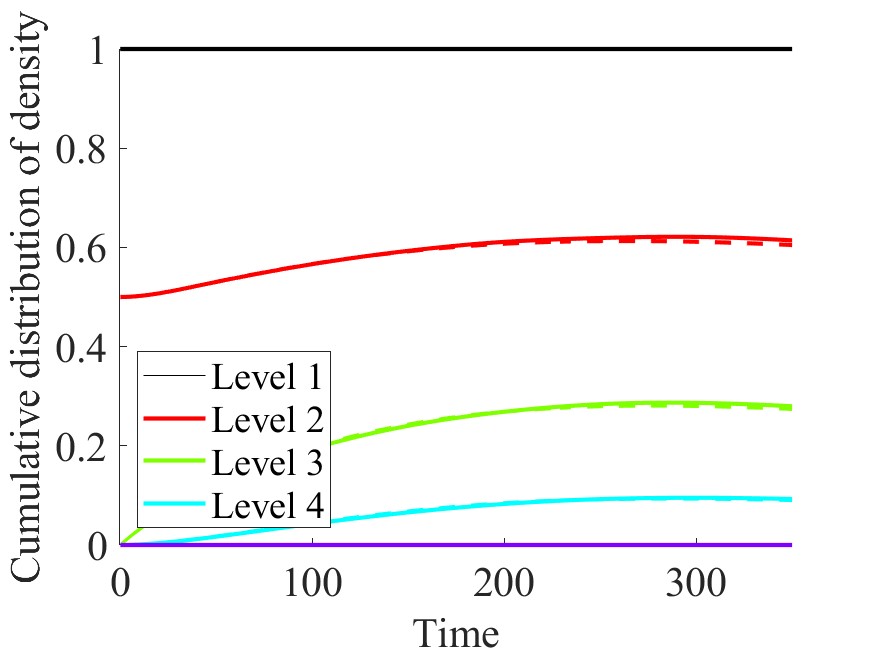}\label{fig_opposing_directions_height_distributions}}
\caption{Opposing groups of vehicles}\label{fig_opposing_directions}
\end{figure}
\Fig \ref{fig_opposing_directions} shows the traffic operations in scenario 1, with opposing streams. \Fig \ref{fig_init_opposing} shows the operations at one flight level. Red bars and (magenta for visibility) arrows represent traffic which needs to move further north (i.e., less south, so to the top of the figure). Blue bars and (cyan for visibility) arrows represent traffic that want to south (down in the figure). The figure shows one same flight level. Indeed, the traffic streams in different classes moves around each other. The red UAVs were initialized at the left of the figure, and the blue ones at the right, with overlap. The red ones are indeed pushed further to the left and the blue ones further to the right, in order to pass each other. Let's also analyze the flight levels, \fig \ref{fig_opposing_directions_height_distributions}. For simplicity, we opted for a 4 level situation, where we initialized both classes in the upper 2 levels (level 1 and 2). Over time, the density in these levels gradually decreased because other UAVs push them away. Note that level 2 remains relatively constant at first because while UAVs are pushed to level 3, other UAVs from level 1 enter level 2. After some time, the density in these levels becomes lower and the density in each of the levels can remain more constant. The distributions are the same for both types of UAVs.

\begin{figure}
\subfigure[Start]{\includegraphics[width=\halvepagina]{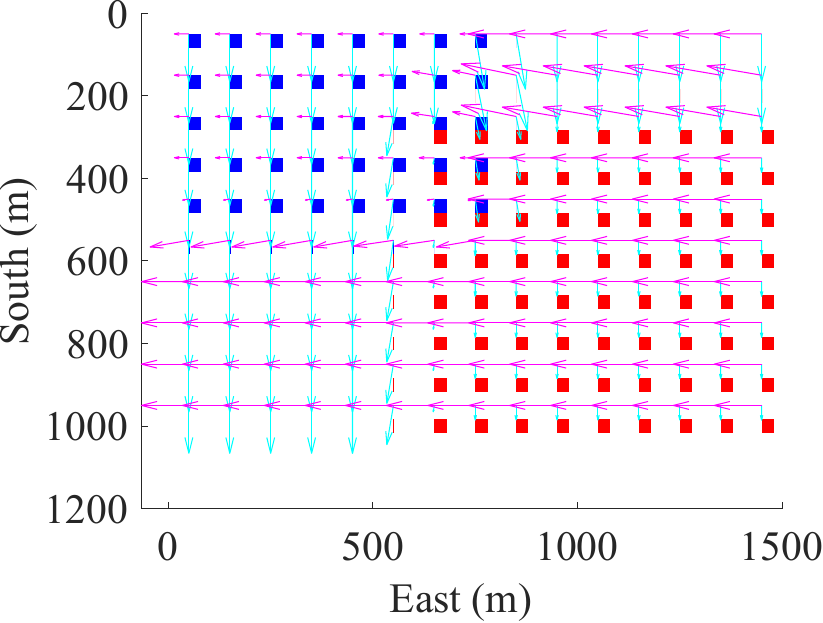}\label{fig_crossing_0}}
\subfigure[In flight]{\includegraphics[width=\halvepagina]{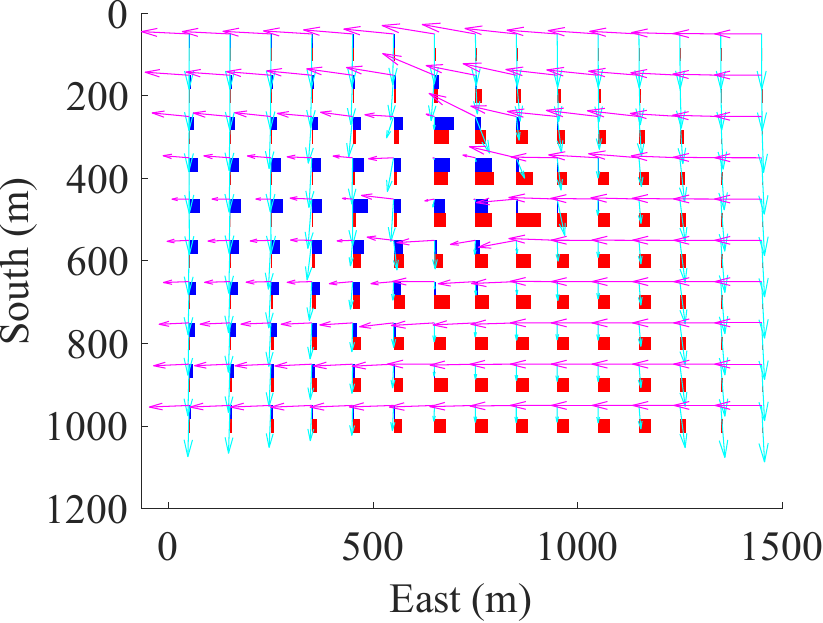}\label{fig_crossing_1000}}
\subfigure[Vertical distribution]{\includegraphics[width=\halvepagina]{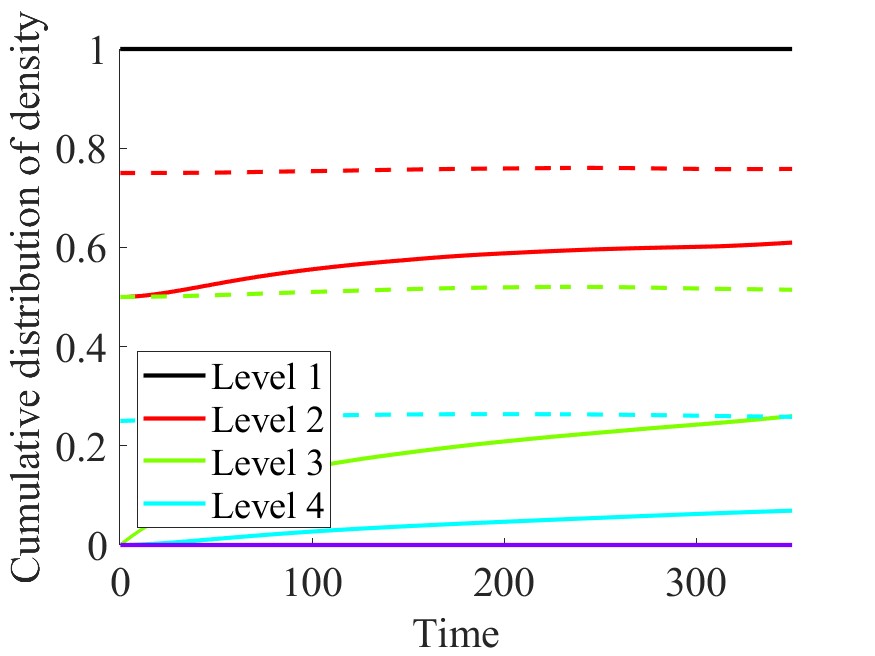}\label{fig_crossing_height_distributions}}
\caption{Traffic operations of the crossing scenario}\label{fig_crossing}
\end{figure}

For the second scenario, we see a pattern of UAVs trying to evade each other, see \fig \ref{fig_crossing}. The figure shows two classes of UAV heading into each other under a 90 degree angle (\fig \ref{fig_crossing_0}). The directional evasive manouver wihtin flight level can be derived from the speed figure. The paths of the different classes of UAV collide. At start, the right-to-left UAVs avoid the blue ones by taking a path that is lower in the figure (higher X value). After some time, see \fig \ref{fig_crossing_1000}, the blue UAVs moving in the X direction also have occupied those cells. Then the red UAVs swivel around at both sides. The blue UAVs at first avoid the red ones by moving to the left in the figure. Since the length of the group of red UAVs is larger, even after a while (\fig \ref{fig_crossing_1000}), there is no option to go around at the right hand side.

With regard to the altitudes: there is a different number of UAVs and a different distribution at start. The the lines in \fig \ref{fig_opposing_directions_height_distributions} at t=0 show that one class (solid line, being the UAVs moving top to bottom; blue in \fig \ref{fig_crossing_0}) are distributed equally over level 1 and 2. The other class, the dotted lines (ones moving right to left; red in \fig \ref{fig_crossing_0}) are divided over 4 levels. They are more in number as well. The blue ones are diverted more from the initial flight heights than the red ones. This is because the red ones have more ``pushing power'' to move some other away, in this case to other flight levels, and also have a higher inertia as collective due to their number (note that the individual movements are equal). This is conceptually related to the clustering into stripes we found in the microscopic description: the UAVs that move in a different direction are being pushed away, which means that UAVs moving in the same direction cluster together.

\begin{figure}
\subfigure[Start]{\includegraphics[width=\halvepagina]{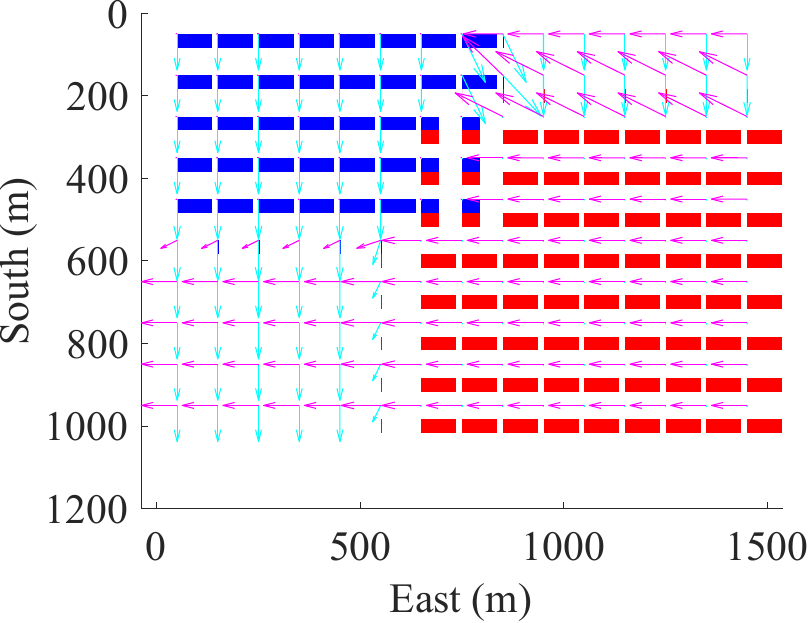}\label{fig_crossing-overcritical_0}}
\subfigure[In flight]{\includegraphics[width=\halvepagina]{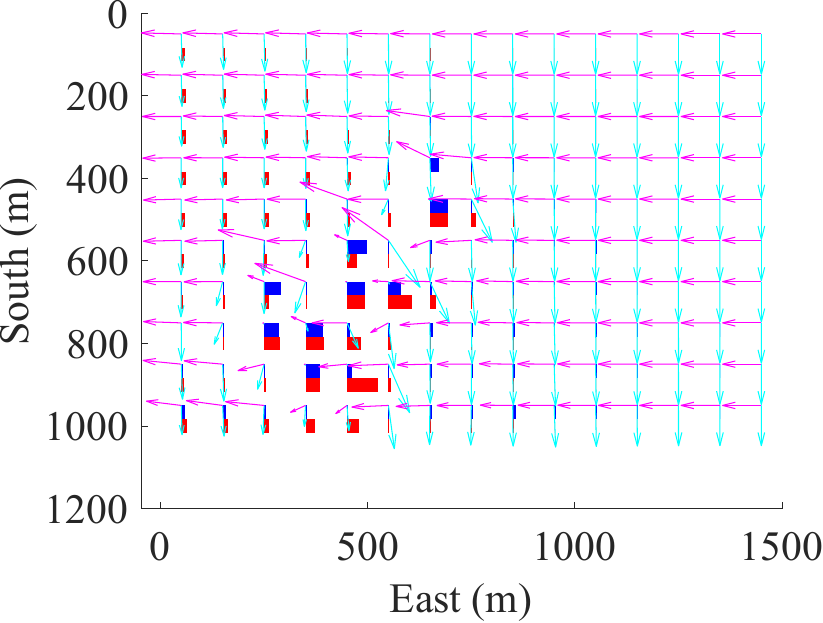}\label{fig_crossing-overcritical_6000}}
\caption{Traffic operations of the crossing scenario with a higher demand}\label{fig_crossing-overcritical}
\end{figure}

We repeat the same scenario with a higher initial demand. This is shown in \fig \ref{fig_crossing-overcritical}. Initially, \fig \ref{fig_crossing-overcritical_0} give already rise to a higher spatial deviation of the paths (the arrows are deviating more from the straight paths than in the scenario with the lower densities). 

We also observe a completely other phenomenon. This crossing scenario gives rise to some cells where density increases and speeds keep decreasing to low values (see e.g., 400m east 700m north). The speeds are low and the speeds of the UAVs direct them around this area. Note that this is based on the macroscopic model, derived from the social force model, and no path planning is applied. Nonetheless, the path planning seems to work quite efficient with most UAVs going around this cell because they are being pushed away from the high density. That does not hold for the UAVs already in the high density cell. However, the high density remains active in this cell for a long period of time, so a cooperative strategy could help to relieve these issues. 

\begin{figure}
\subfigure[In-flight-level operations]{\includegraphics[width=\halvepagina]{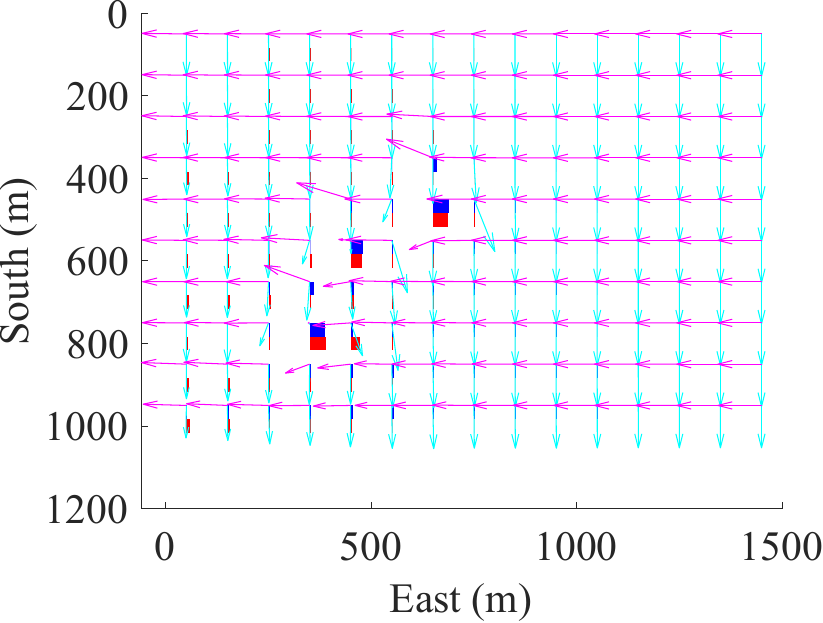}\label{fig_jammed_lambda0}}
\subfigure[Number of UAVs]{\includegraphics[width=\halvepagina]{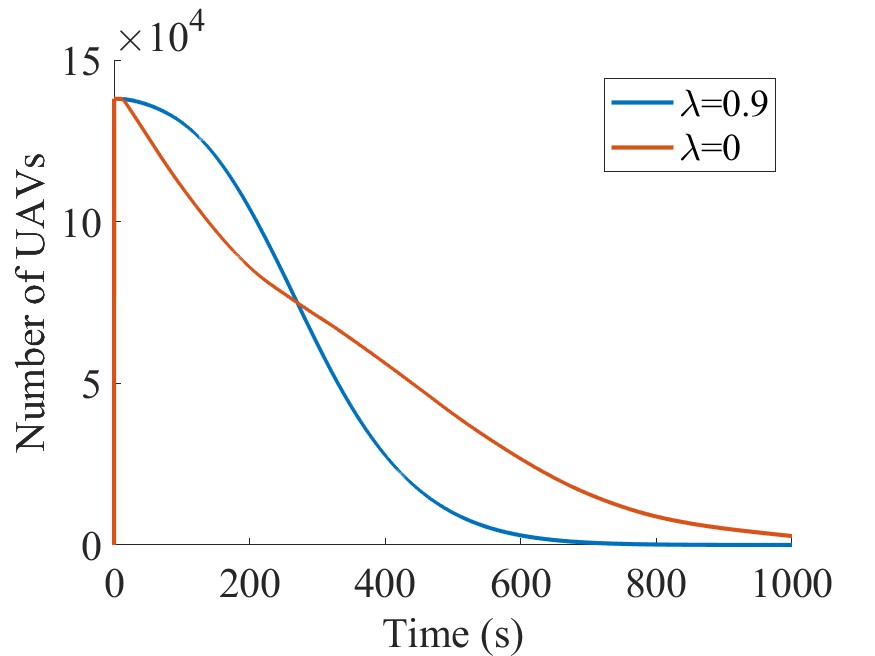}\label{fig_comparison_nr_drones}}
\caption{The effect of $\lambda$}\label{fig_effect_of_lambda}
\end{figure}

The value of $\lambda$ also plays a role here. If UAVs have a stronger repelling action from UAVs with a different direction of speed, there is a lower tendency to end up in a situation with very high density where two classes of UAV compete for the same crossing space. We repeated the same experiments with overcritical crossing flows for situation with no asymmetry ($\lambda$=0). Results are shown in \fig \ref{fig_effect_of_lambda}. Without this extra avoidance of UAVs with different directions, the flows get ``jammed'' more severely, and in several cells (see \fig \ref{fig_jammed_lambda0}. Consequently, traffic is less efficient under these high density conditions. This shows in the arrivals. \Fig \ref{fig_comparison_nr_drones} shows that the number of UAVs in the system remains for $\lambda=0$ and $\lambda=0.9$. At first, the more severe interaction term with higher lambda causes for stronger interaction between the classes and path deviations. But once these are settled, we see a higher arrival rate (line goes down more steeply), matching the time that the bulk of the initial demand would arrive with the slightly modified paths. On the other hand, analyzing the line for $\lambda=0$, we see that the beginning is not affected so much, but also there is not the same high arrival rate as we see for the case $\lambda=0.9$, which is indeed due to the UAVs being blocked internally and only being able to escape this blockage gradually. The breaking of symmetry can hence, also in the macroscopic modeling, increase traffic efficiency in high density conditions. 

\begin{figure}
\subfigure[Only supply restriction]{\includegraphics[width=\halvepagina]{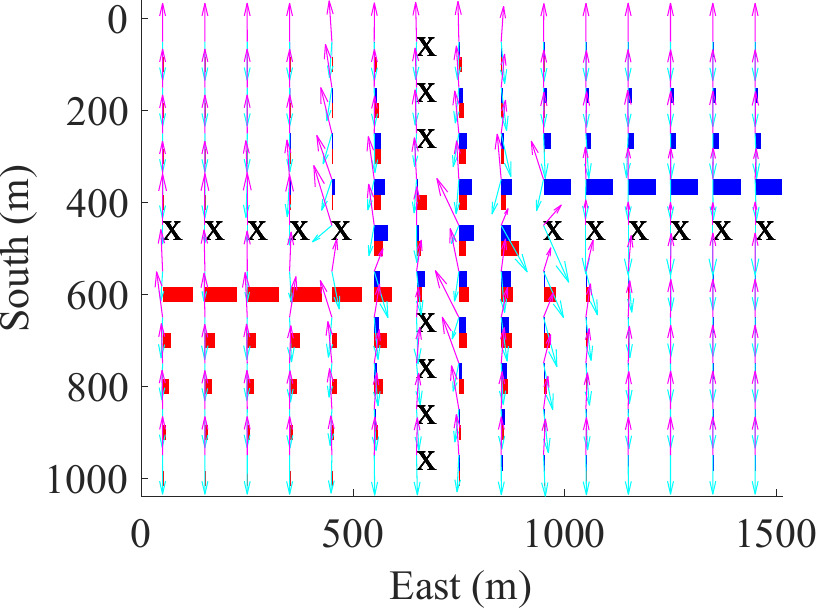}\label{fig_bottleneck_initial}}
\subfigure[Supply restriction and repelling density]{\includegraphics[width=\halvepagina]{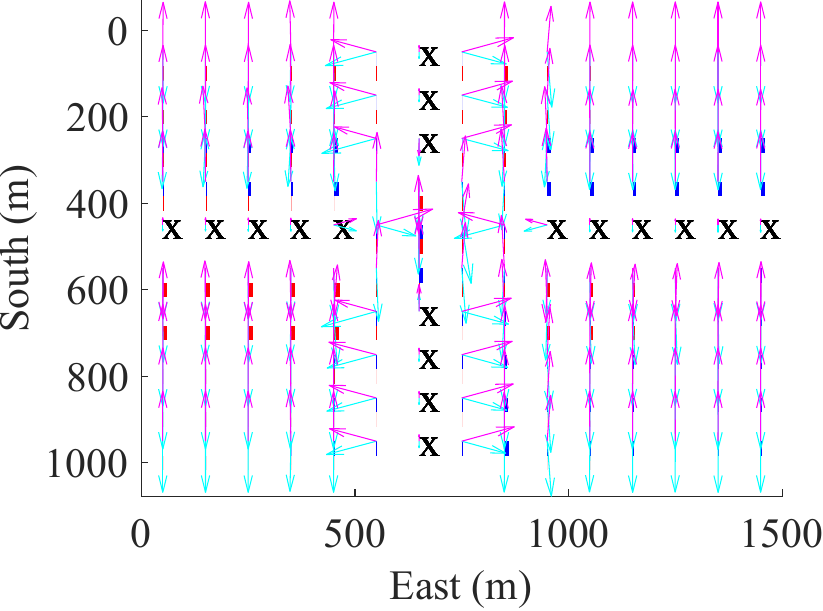}\label{fig_bottleneck_high_repulsion}}\\
\subfigure[Flight levels; dashed lines are only supply restriction and solid lines also include artifical density increase]{\includegraphics[width=\halvepagina]{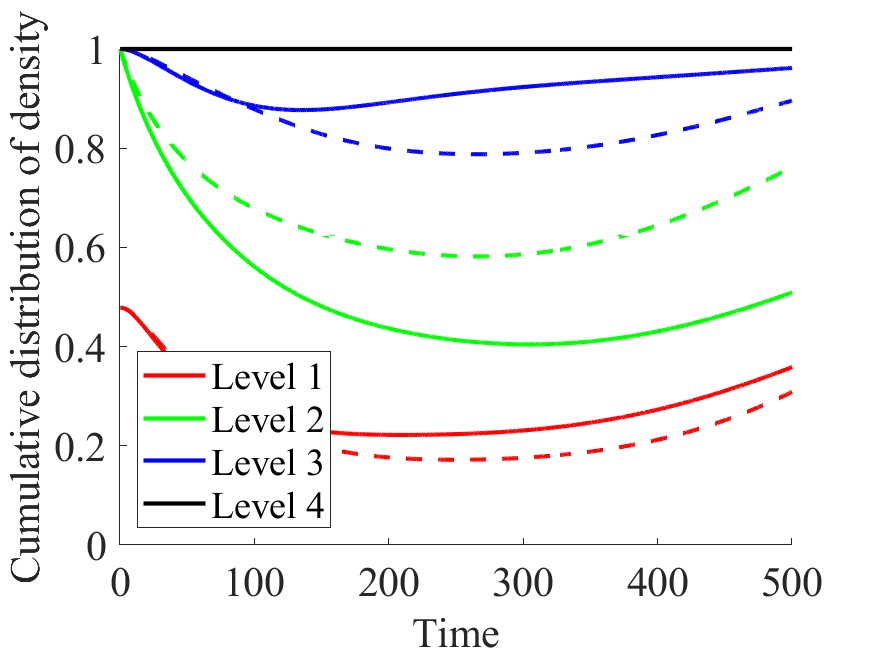}\label{fig_bottleneckFL}}
\end{figure}

To analyze the effect of a bottleneck, we model a case with two blocked rows of buildings at the lowest layer, indicated by crosses in \fig \ref{fig_bottleneck_initial} (scenario 5 in \tab \ref{tab_macroscopic_scenarios}; level 2 being shown). If only the supply is restricted, UAVs can fly in the cells immediately next to the buildings. If also the cells themselves receive artificial density (\fig \ref{fig_bottleneck_high_repulsion}), the blocked cells repel the UAVs, the velocity vectors push the UAVs outwards, and the cells next to the blocked cells have very low densities too. For the UAVs there are two ways for the UAVS to reach the destination: (i) through the middle in between the blockings or (ii) in another, higher layer. In both cases, there are UAVs at the side (e.g. at 100m east and 600m south, heading north) that do not have the immediate force to the middle and cannot easily pass. In the case of artificial density, there is more repelling to lower levels, so more UAVs will take that path. Hence, UAVs will change flight levels. Note that the most important change comes once they have increased the flight level to level 2. In case the blocked cells also repel, the UAVs are pushed further towards level 3. Hence, the number of UAVs in level 2 much lower if the repelling force from artificial density is included. That is visible in \fig \ref{fig_bottleneckFL} because vertical distance between the level 2 and level 3 (the amount of UAVs in level 2) is much bigger in case of the supply-restriction only (solid line), and in the dashed line, this distance is small, hence there are few UAVs in level 2. Ideally and more realistically, the repelling force is not so extreme near the point where UAVs can pass the bottleneck, and has more repelling power at larger distances. This also means that the repelling will have have some kind of gradient towards a feasible route, which is also desired to avoid ``jammed'' UAVs. Model wise, this could be resolved by having a radius for a repelling density which is in the same order as the blocked length. That means different value of the repelling radius: smaller for UAVs, and larger for buildings.  Combining this for a wall then would lead to less repelling force near the edges because -- in the view of the UAV interacting with the wall -- near the edges a smaller fraction of the radius is interacting with the wall. 


\section{Conclusions and outlook}\label{sec_conclusions}
In this abstract we have explored the interactions for UAVs operating in 3D without a form of centralized control and without a predictive path planning. We listed the criteria which interactions in 3D should (ideally) satisfy, as well as criteria for effectiveness of interactions.

Microscopically, we have explored the social force model, and introduced an impact factor to account for asymmetric interactions. This way, UAVs can interact more effectively with other UAVs.  The model was found to produce efficient trajectories. The UAVs started deviating their path relatively far upstream of the interaction point. That means that the UAVs can pass each other relatively easily. One negative side of (symmetric) interactions is that both UAVs can push each other to the side. The model with asymmetric interactions improved in this respect. 

The paper also how dynamics of large numbers of UAVs can be described via a macroscopic model, and what dynamic patterns emerge there. We innovated a modeling framework incorporating a numerical scheme to solve multi-class, multi-dimensional and multi-directional traffic in one macroscopic modeling framework. This model was run showing the macroscopic patterns of UAV traffic.

In both scales, we find that UAVs cluster, and that it helps to breaking symmetry of interactions between UAVs. If UAVs are moving as a group, they are less susceptible to influences of UAVs moving in another direction, which causes them to move together. As group, they can create space for their group to pass due to their collective ``force'' on other UAVs. This is present in both the microscopic and macroscopic results. 

In the macroscopic description, areas could be made inaccessible by limiting the supply to zero. Moreover, a repulsive force could be added by artificially increasing the density to repel UAVs. Whereas the added density could get a more effective repulsion from the inaccessible areas, it also proved to be not very subtle. The repelling forces pushes UAVs away from cells directly neighboring the inaccessible cells as well. An alternative implementation might include a zero supply as well as a repulsive action, but \emph{with a different strength and radius $R$}. A longer radius of repulsion, comparable to the size of the object, would start to repel earlier and also gradually decrease towards an opening, thereby creating the force towards a feasible path. 

Looking ahead in time for a situation can avoid getting stuck, and can help the flow in future conditions. There is a natural tendency to segregate UAVs per class (i.e., UAVs with a similar direction of speed cluster together). This reduces negative interactions and also can improve traffic flows. The macroscopic descriptions comparable to that show that less interactions with UAVs moving in the same direction would smooth UAV traffic and make it more efficient. 

For UAV traffic, contrary to pedestrian and car traffic, there is no aim to calibrate or validate the models in order to get the correct, observed behavior. Instead, the insights and models can be used to actually design safe and efficient interaction between UAVs, among others by properly designing the interaction rules. This paper linked instantaneous microscopic interactions to macroscopic patterns. Future work should address the macroscopic effects of other types of interactions, notably interactions that used planned path (responding to future interactions, i.e. not instantaneous), and flight rules that take spatial anisotropic rules into account (e.g., UAVs following 3D roads with different properties).

\bibliographystyle{apalike} 
\bibliography{bibs_arXiv}
\end{document}